\documentclass[letterpaper,twocolumn,prx,aps,showpacs,superscriptaddress,floatfix]{revtex4-1}
\usepackage[latin1]{inputenc}
\usepackage{bm}
\usepackage[usenames]{color}
\usepackage{multirow}
\usepackage{amssymb}
\usepackage{amsbsy}
\usepackage{amsmath}
\usepackage{mathtools}
\usepackage{stmaryrd}
\usepackage{graphicx}
\usepackage{epsfig}
\usepackage{placeins}
\usepackage{ulem}
\usepackage{braket}
\usepackage[colorlinks,linkcolor=blue,citecolor=blue,urlcolor=blue]{hyperref}
\usepackage{soul}

\makeatletter
\begin{document}
\title{Sudden removal of a static force in a disordered system: \\
Induced dynamics, thermalization, and transport}

\author{Jonas Richter}
\email{jonasrichter@uos.de}
\affiliation{Department of Physics, University of Osnabr\"uck, D-49069 Osnabr\"uck, Germany}

\author{Jacek Herbrych}
\email{jherbryc@utk.edu}
\affiliation{Department of Physics and Astronomy, The University of Tennessee, Knoxville, Tennessee 37996, USA}
\affiliation{Materials Science and Technology Division, Oak Ridge National Laboratory, Oak Ridge, Tennessee 37831, USA}

\author{Robin Steinigeweg}
\email{rsteinig@uos.de}
\affiliation{Department of Physics, University of Osnabr\"uck, D-49069 Osnabr\"uck, Germany}

\date{\today}

\begin{abstract}
We study the real-time dynamics of local occupation numbers in a one-dimensional 
model of spinless fermions with a random on-site potential for a certain class 
of initial states. The latter are thermal (mixed or pure) states of 
the model in the presence of an additional static force, but become 
non-equilibrium states after a sudden removal of this static force. For this 
class and high temperatures, we show that the induced dynamics is given by a 
single correlation function at equilibrium, independent of the initial 
expectation values being prepared close to equilibrium (by a weak static force) 
or far away from equilibrium (by a strong static force). Remarkably, this type 
of universality holds true in both, the ergodic phase and the many-body 
localized regime. Moreover, it does not depend on the specific choice of a unit 
cell for the local density. We particularly discuss two important consequences. 
First, the long-time expectation value of the local density is uniquely 
determined by the fluctuations of its diagonal matrix elements in the energy 
eigenbasis. Thus, the validity of the eigenstate thermalization hypothesis is 
not only a sufficient but also a necessary condition for thermalization. Second, 
the real-time broadening of density profiles is always given by the current 
autocorrelation function at equilibrium via a generalized Einstein relation. 
In the context of transport, we discuss the influence of disorder for large 
particle-particle interactions, where normal diffusion is known to occur in the 
disorder-free case. Our results suggest that normal diffusion is stable against 
weak disorder, while they are consistent with anomalous diffusion for stronger
disorder below the localization transition. Particularly, for weak disorder, Gaussian density profiles can be 
observed for single disorder realizations, which we demonstrate for finite lattices up to $31$ sites.
\end{abstract}

\maketitle


\section{Introduction}

Statistical mechanics provides a universal concept to describe the properties
of many-body quantum systems at equilibrium, and a microscopic treatment of the
exponentially many degrees of freedom is replaced in favor of associating the
system with a few macroscopic parameters like energy or temperature. Out of
equilibrium, however, such a universal concept is absent. This fact is not
least due to the multitude of different non-equilibrium scenarios, e.g., the
system can be driven by time-dependent protocols \cite{Eckardt2005, 
Lazarides2014}, or it can be in contact with heat baths or particle reservoirs
at unequal temperatures or chemical potentials \cite{Monasterio2005, 
Michel2008, Znidaric2011}, just to name a few possibilities.

On the contrary, for quantum systems in strict isolation, a non-equilibrium
situation can only be induced by the preparation of suitable initial states,  
e.g., by means of a quench \cite{Essler2014, Karrasch2014, Essler2016}. These 
initial states can be mixed or pure, entangled or non-entangled, and their 
properties might be of essential importance for the subsequent relaxation 
process \cite{Reimann2016, Richter2018, James2018}. In this context, the 
intriguing question arises whether the system will eventually reach thermal 
equilibrium under its own unitary dynamics governed by the Schr\"odinger 
equation. This fundamental question has attracted a lot of interest in recent 
years \cite{Polkovnikov2011, Eisert2015, Gogolin2016, Dallesio2016}, and it has 
also profited from the interplay between theory and experiment. On the one hand, 
cold atomic gases and trapped ions provide ideal testbeds to experimentally 
study almost perfectly isolated systems in a controlled manner \cite{Bloch2005, 
Langen2015, Trotzky2012, Blatt2012}. On the other hand, emergent theoretical 
concepts such as the eigenstate thermalization hypothesis (ETH) 
\cite{deutsch1991, srednicki1994, rigol2005} and the typicality of pure quantum 
states \cite{Gemmer2004, Popescu2006, Goldstein2006, Reimann2007}, as well as 
the development of powerful numerical techniques \cite{Schollwock2011}, have 
deepened our understanding of equilibration in closed quantum systems.

While it is commonly expected that generic quantum many-body systems fulfill 
the ETH \cite{Mondaini2017}, there are also exceptions, of course. An obvious 
class of such counterexamples is given by integrable quantum systems, where 
thermalization to standard statistical ensembles is prevented by a macroscopic 
number of (quasi-local) conserved quantities \cite{Zotos1997, Prosen2013}. 
Nonetheless, a concise description of such systems in terms of so-called 
generalized Gibbs ensembles still remains possible \cite{Rigol2007, Vidmar2016,  
Ilievski2015}. Another class of models which fail to thermalize are disordered 
quantum systems, where many-body localization (MBL) can occur for sufficiently 
strong disorder \cite{Basko2006, Nandkishore2015, Altman2015}. While the mere 
existence of the MBL phase has been confirmed both, numerically and 
analytically for certain models \cite{Oganesyan2007, Berkelbach2010, 
Imbrie2016}, and its experimental realization has seen substantial progress 
recently \cite{Schreiber2015, Smith2016, Abanin2018}, a full understanding 
of disordered many-body quantum systems out of equilibrium continues to be a 
challenge.

In this paper, we study the real-time dynamics of local occupation numbers in a 
one-dimensional model of spinless fermions with a random on-site potential for 
a certain class of initial states. The latter are thermal (mixed or 
pure) states of the model in the presence of an additional static force, but 
become non-equilibrium states after a sudden removal of this static force. For 
this class and high temperatures, we show that the induced dynamics is given by 
a single correlation function at equilibrium, independent of the initial 
expectation values being prepared close to equilibrium (by a weak static 
force) or far away from equilibrium (by a strong static force). Remarkably, 
this type of universality holds true in both, the ergodic phase and the 
many-body localized regime. Moreover, it does not depend on the specific choice 
of a unit cell for the local density.

While our model is certainly different, these results are also relevant to 
recent experiments which report on the occurrence of universal dynamics far 
from equilibrium during the relaxation of an isolated one-dimensional Bose gas 
\cite{Prufer2018, Erne2018}. Moreover, we discuss two important consequences. 
First, the long-time expectation value of the local density is uniquely 
determined by the fluctuations of its diagonal matrix elements in the energy 
eigenbasis. Thus, the validity of the eigenstate thermalization hypothesis is 
not only a sufficient but also a necessary condition for thermalization. 
Second, the real-time broadening of density profiles is always given by the 
current autocorrelation function at equilibrium via a generalized Einstein 
relation. In the context of transport, we discuss the influence of disorder for 
large particle-particle interactions, where normal diffusion is known to occur 
in the disorder-free case \cite{Prelovsek2004, Znidaric2011, Steinigeweg2011, 
Karrasch2014, Steinigeweg2017, Ljubotina2017}. Our results suggest that normal 
diffusion is stable against weak disorder, while they are consistent with 
anomalous diffusion for stronger disorder.

This paper is structured as follows: We introduce the model in Sec.\ 
\ref{sec:model} and the non-equilibrium setup in Sec.\ \ref{sec:ned}. In Sec.\ 
\ref{sec:results} we turn to our results, where we start with the occurrence of
universal dynamics in Sec.\ \ref{Sec:Universal} and continue with its 
consequences for thermalization in Sec.\ \ref{sec:thermalization} and transport 
in Sec.\ \ref{Sec:Broad}. We summarize and conclude in Sec.\ 
\ref{sec:conclusion}.


\section{Model} \label{sec:model}

We study a one-dimensional model of spinless fermions with a random on-site 
potential and periodic boundary conditions (PBC), described by the Hamiltonian 
\begin{align} \label{Eq::Ham}
\mathcal{H} & = J\sum_{l=1}^L \biggl [ \tfrac{1}{2}(c_l^\dagger c_{l+1} + 
\text{H.c.}) + \Delta \left(n_l - \tfrac{1}{2}\right) \hspace{-0.1cm} 
\left(n_{l+1} -  \tfrac{1}{2} \right) \nonumber\\
& + \mu_l \left(n_l - \tfrac{1}{2} \right) \biggr ] \ ,  
\end{align}
where $c_l^{\dagger}$ ($c_l$) creates (annihilates) a spinless fermion 
at lattice site $l$, $n_l = c_l^\dagger c_l$ is the occupation number, and $L$ 
is the number of sites. $J$ sets the energy scale and $\Delta$ is the strength 
of the nearest-neighbor interaction. The potentials $\mu_l$ are randomly drawn 
from a uniform distribution in the interval $\mu_l \in [-W, W]$. Note that, due 
to the Jordan-Wigner transformation, $\mathcal{H}$ is identical to the 
spin-$1/2$ XXZ chain with a random magnetic field. Note further that 
$\mathcal{H}$ is integrable for $W = 0$ in terms of the Bethe Ansatz, with the 
energy current being exactly conserved \cite{Klumper2002}. Since $\mathcal{H}$ 
conserves the total charge, i.e., $[\mathcal{H}, \sum_l n_l] = 0$, the particle 
current $j$ is well-defined via a lattice continuity equation and takes on the 
form
\begin{equation} \label{Eq::Current}
j = \frac{J}{2}\sum_{l=1}^L (i c_l^\dagger c_{l+1} + \text{H.c.})\ .
\end{equation}
We have $[\mathcal{H}, j] = 0$ in the limiting case $\Delta = W = 0$ only, 
while generally $[\mathcal{H}, j] \neq 0$ for any other choice of $\Delta$ or 
$W$ (although it is known that $j$ is partially conserved for $\Delta < 1$ and 
$W = 0$ \cite{Zotos1999, HeidrichMeisner2003, Prosen2013}). 

The Hamiltonian \eqref{Eq::Ham} (or its spin-chain counterpart) is 
an archetypal model \cite{Znidaric2008, Pal2010, Bera2015, Herrera2015, 
Luitz2015, Hauschild2016, Steinigeweg2016, Mierzejewski2016, Prelovsek2017, 
Schmidtke2017} to study the disorder-driven transition between an ergodic 
regime ($W < W_\text{c}$) and an MBL phase ($W > W_\text{c}$), where $W_c$ 
denotes some critical disorder value. For the mostly studied case $\Delta = 1$, 
$W_\text{c}$ has been suggested to be approximately $W_\text{c} \approx 3.5$, 
although the numerical analysis is a severe challenge. While disordered systems 
certainly feature various fascinating properties (see e.g.\ Refs.\ 
\cite{Bardarson2012, Huse2014, Khemani2015, Vasseur2016}), let us here focus on 
only two aspects: ETH and transport. For weak disorder $W < W_\text{c}$, the 
ETH is expected to hold, and the system thermalizes at long times. In contrast, 
for strong disorder $W > W_\text{c}$, the ETH is not fulfilled and the system 
does not thermalize. Recently, there has also been increased interest in 
exploring the ergodic side of the MBL transition and transport 
\cite{Luitz2017}. In this regime, Griffiths effects, i.e., rare events, might 
facilitate the possibility of anomalous transport and subdiffusion 
\cite{Agarwal2015, Gopalakrishnan2015, BarLev2015, Luitz2016, Khait2016, 
Prelovsek2017_2}. In this paper, we will also discuss this issue.


%
\begin{figure}[tb]
\centering
\includegraphics[width=1\columnwidth]{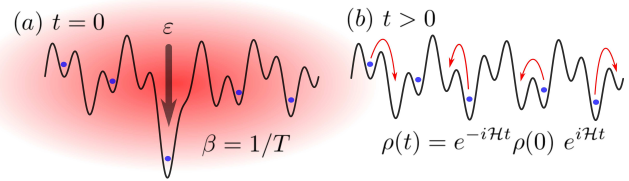}
\caption{(Color online) Sketch of the setup. (a) At time $t = 0$, the disordered 
system is still in contact with a heat bath at inverse temperature $\beta = 
1/T$ and a static force leads to an additional potential in the middle of the 
chain. (b) At times $t > 0$, heat bath and static force are both removed and 
the system evolves unitarily according to the isolated Hamiltonian 
$\mathcal{H}$. This setup might be seen as a type of quantum quench as well.}
\label{Fig1}
\end{figure}
\section{Non-equilibrium setup} \label{sec:ned}

\subsection{Initial states}

In this paper, we investigate the dynamical expectation values of local 
occupation numbers 
\begin{equation}
\langle n_l(t) \rangle = \bra{\psi(t)} n_l \ket{\psi(t)}\ , 
\end{equation}
with $\ket{\psi(t)} = e^{-i \mathcal{H} t} \ket{\psi(0)}$, where 
$\ket{\psi(0)}$ is a suitably prepared non-equilibrium pure state,
\begin{equation} \label{Eq::State1}
\ket{\psi(0)} = \frac{\sqrt{\rho} \ket{\varphi}}{\sqrt{\bra{\varphi} \rho 
\ket{\varphi}}}\ .
\end{equation}
Here, the pure reference state $\ket{\varphi}$ is prepared according to 
the unitary invariant Haar measure, i.e.,
\begin{equation}
\ket{\varphi} = \sum_k c_k \ket{\varphi_k} \, , \label{phi}
\end{equation}
where the $c_k$ are complex numbers drawn at random from a Gaussian 
distribution with mean zero. The $\ket{\varphi_k}$ denote orthogonal basis 
states of the Hilbert space, e.g., the common eigenbasis of all $n_l$.
If not stated otherwise, we always consider the full Hilbert space, i.e., all sectors of fixed charge.

The operator $\rho$ in Eq.\ \eqref{Eq::State1} can be thought of as a density 
matrix resulting from the following physical scenario: Consider a quantum 
system which is (weakly) coupled to a (macroscopically large) heat bath at 
inverse temperature $\beta = 1/T$. Moreover, let the system be affected by a 
external static force, which (i) gives rise to an additional potential of 
strength $\varepsilon$ and (ii) is spatially restricted to the center of the 
lattice, i.e., $L/2$. (Note that, due to PBC, this particular choice is arbitrary.)
Then, at equilibrium, this situation is described by 
the density matrix \cite{kubo1991, Bartsch2017, Richter2017, Richter2018_2}
\begin{equation} \label{Eq::DensityMat}
\rho = e^{-\beta(\mathcal{H}-\varepsilon n_{L/2})} / \mathcal{Z}\ , 
\end{equation}
where $\mathcal{Z} = \text{Tr}[e^{-\beta(\mathcal{H}-\varepsilon n_{L/2})}]$ 
denotes the partition function. By removing both, the heat bath as well as the 
external force, one can induce a non-equilibrium situation (see Fig.\ 
\ref{Fig1}), where $\rho$ is an out-of-equilibrium state of the remaining Hamiltonian 
$\mathcal{H}$ and evolves according to the von-Neumann equation, $\rho(t) = 
e^{-i\mathcal{H}t} \rho(0) e^{i\mathcal{H}t}$.

In the sense of typicality \cite{Bartsch2009, Sugiura2013, Elsayed2013, 
Steinigeweg2014_2, iitaka2003, Reimann2018}, the pure states $\ket{\psi(0)}$ in 
Eq.\ \eqref{Eq::State1} represent a whole ensemble of valid initial states from 
the Hilbert space, which most likely mimic the density matrix in Eq.\ 
\eqref{Eq::DensityMat}. In particular, we can write
\begin{equation} \label{Eq::Typicality}
\text{Tr}[\rho(t) n_l] = \bra{\psi(t)} n_l \ket{\psi(t)} + f(\ket{\varphi})\ , 
\end{equation}
where the statistical error $f(\ket{\varphi})$ scales as 
$f(\ket{\varphi})\propto 1/\sqrt{d_\text{eff}}$ with the effective Hilbert-space dimension $d_\text{eff}$. 
Specifically, $d_\text{eff} = \mathcal{Z}/e^{-\beta E_0}$ is a partition function and $E_0$ is the ground-state energy of $\mathcal{H} - \varepsilon n_{L/2}$. 
Thus, $f(\ket{\varphi})$ vanishes exponentially fast for increasing 
system size. Particularly, for $\beta \rightarrow 0$, $d_\text{eff} = 2^L$ and $f(\ket{\varphi})$ can be neglected 
for medium-sized systems already.

Let us discuss some of the properties of this class of initial states. On the 
one hand, for $\varepsilon \rightarrow 0$, one naturally finds $\rho 
\rightarrow \rho_\text{eq}$, where $\rho_\text{eq} = \exp(-\beta 
\mathcal{H})/\mathcal{Z}_\text{eq}$ denotes the equilibrium density matrix of 
the canonical ensemble with $\mathcal{Z}_\text{eq} = \text{Tr}[\exp(-\beta 
\mathcal{H})]$. Consequently, for all $n_l$, we find the initial expectation 
value 
\begin{equation}
\lim_{\varepsilon \rightarrow 0} \bra{\psi(0)} n_l \ket{\psi(0)} = n_\text{eq}\ ,  
\end{equation}
with $n_\text{eq} = \text{Tr}[\rho_\text{eq} n_l]$. On the other hand for $\varepsilon \rightarrow 
\infty$, $\rho$ acts as a projection onto the eigenstates of $n_{L/2}$ with the 
largest eigenvalue $n_\text{max} = 1$. For the particular case of $n_{L/2}$, we 
therefore have
\begin{equation}
\lim_{\varepsilon \rightarrow \infty} \bra{\psi(0)} n_{L/2} \ket{\psi(0)} = 
n_\text{max}\ .   
\end{equation}
Thus, by varying the strength of the external force from small to large 
$\varepsilon$, one can prepare initial states which are close to equilibrium, 
i.e.,
$\langle n_{L/2}(0) \rangle \approx n_\text{eq}$, or in contrast also 
states which are maximally far from equilibrium, i.e., $\langle n_{L/2}(0) 
\rangle \approx n_\text{max}$ \cite{Richter2017, Richter2018_2}.

It is instructive to discuss the regime of small perturbations in more detail. 
Here, we can expect from linear response theory that \cite{kubo1991}
\begin{equation} \label{Eq::Linear}
\langle n_l(t) \rangle = n_\text{eq} + \varepsilon \, \chi_{L/2,l}(t) \, ,
\end{equation}
where $\chi_{L/2,l}(t) = \beta(\Delta n_{L/2}; n_l(t))$ is given by a Kubo 
scalar product
\begin{equation}\label{Eq::Chi}
\chi_{L/2,l}(t) = \int_0^\beta \text{d}\lambda\ \text{Tr}[e^{-\beta \mathcal{H}} 
\Delta n_{L/2}(-i\lambda) n_l(t)]\ , 
\end{equation}
with $\Delta n_{L/2} = n_{L/2} - n_\text{eq}$ and $\Delta n_{L/2}(-i\lambda) =  
e^{\lambda \mathcal{H}} \Delta n_{L/2} e^{-\lambda \mathcal{H}}$. For large 
$\varepsilon$, i.e., outside the linear response regime, the linear relationship 
in Eq.\ \eqref{Eq::Linear} is generally expected to break down. Thus, it is an 
important question how the dynamics of $\langle n_l(t) \rangle$ evolves for 
initial states far from equilibrium.

While it is in principle possible to study this question for arbitrary $\beta$, 
we here want to focus on the regime of high temperatures. Specifically, in the  
limit $\beta \to 0$ but finite $\beta \varepsilon$, we have in good 
approximation $\rho \propto e^{\beta \varepsilon n_{L/2}}$, i.e., the 
Hamiltonian is irrelevant for the initial state $\ket{\psi(0)}$. Note that the 
subsequent dynamics, on the contrary, significantly depends on $\mathcal{H}$. 
In the $\beta\to 0$ limit, Eq.\ \eqref{Eq::Chi} can be 
simplified to
\begin{equation}
\chi_{L/2,l}(t) \approx \beta \Big ( \frac{\text{Tr}[n_{L/2} n_l(t)]}{2^L} - 
n_\text{eq}^2 \Big ) \, , 
\end{equation}
and for time $t = 0$ we have $\chi_{L/2,l}(0) \approx \chi_{L/2,l}(0) 
\delta_{L/2,l}$, where $\delta_{L/2,l}$ denotes the Kronecker $\delta$. Loosely 
speaking, the external force remains unnoticed on lattice sites $l \neq L/2$ at 
high temperatures. Consequently, the initial state $\ket{\psi(0)}$ realizes an 
initial density profile with a $\delta$ peak on top of a homogeneous 
many-particle background, 
\begin{equation}\label{Eq::Profile}
\langle n_l(0) \rangle =
\begin{cases} 
{\cal N}_0 > n_\text{eq},\ &l = L/2 \\
 n_\text{eq},\ &l \neq L/2
\end{cases}\ , 
\end{equation}
where the size of the $\delta$ peak ${\cal N}_0= \bra{\psi(0)} n_{L/2} 
\ket{\psi(0)}$ depends on the strength of the perturbation 
$\varepsilon$, as discussed above. Note that Eq.\ \eqref{Eq::Profile} holds for 
arbitrary $\varepsilon > 0$ and also $W > 0$ (if one averages over suitably many 
instances of disorder), but it will likely break down if temperature is not 
high enough. 

In the remainder of this paper, we will discuss the relaxation dynamics of the 
density profiles given in Eq.\ \eqref{Eq::Profile}. Specifically, we will 
discuss the influence of $\ket{\psi(0)}$ being close to or far away from 
equilibrium, i.e., the influence of the initial peak height ${\cal N}_0$. 
Furthermore, we will shed light on the role of the ETH for the long-time 
behavior of $\langle n_l(t) \rangle$.


\subsection{Pure-state propagation and averaging}

In order to evaluate the expectation value $\langle n_l(t) \rangle$, we here  
rely on the typicality relation in Eq.\ \eqref{Eq::Typicality}. This pure-state 
approach has the main advantage that the action of the exponentials 
$e^{-i\mathcal{H}t}$ and $e^{-\beta(\mathcal{H} -\varepsilon n_{L/2})}$ can be 
efficiently evaluated by a forward propagation in real or imaginary time, 
respectively. While there exist various sophisticated methods such as Trotter 
decompositions \cite{DeRaedt2006}, Chebyshev expansions \cite{Dobrovitski2003, 
Weisse2006}, or Krylov subspace techniques \cite{Varma2017}, we here apply a 
fourth-order Runge-Kutta scheme, where the discrete time step is always chosen 
sufficiently short to ensure small numerical errors \cite{Elsayed2013, 
Steinigeweg2014_2, Herbrych2016}. Thus, no exact diagonalization is needed and, 
since the involved operators also have a sparse matrix representation, 
matrix-vector multiplications can be implemented relatively memory-efficient 
\cite{Wietek2018}.

Moreover, let us reiterate that the statistical error $f(\ket{\varphi})$ in
Eq.\ \eqref{Eq::Typicality} for $\beta \approx 0$ can be neglected for all  
system sizes studied here. Therefore, it is completely sufficient to calculate 
all expectation values from one single state, i.e., only one set of random 
coefficients $c_k$, cf.\ Eq.\ \eqref{Eq::State1} and below. It should be noted, 
however, that since our model \eqref{Eq::Ham} contains random on-site 
potentials $\mu_l$, all expectation values will naturally depend on the specific 
realization of these $\mu_l$. Hence, we perform an averaging over $N$ such 
instances of random configurations,
\begin{equation}
\overline{\langle n_l(t) \rangle} = \frac{1}{N} 
\sum_{i=1}^N \langle n_l(t) \rangle_i \, .
\end{equation}
In this paper, we routinely choose $N = 300$, which turns out to be sufficiently 
large to ensure reliable results. As an illustration, the standard deviation
\begin{equation}\label{Eq::Error}
\Delta \langle n_l(t) \rangle = \sqrt{\overline{\Big [ \langle n_l(t) 
\rangle \Big]^2} - 
\Big [\overline{\langle n_l(t) \rangle} \Big ]^2}
\end{equation}
is later shown in Fig.\ \ref{Fig::Fig3} (c) for disorder strengths $W = 1$ and  
$W = 4$. 
Although one finds that $\Delta \langle n_l(t) \rangle$ becomes 
significantly larger for increasing $W$, the error of the average $\Delta 
\langle n_l(t) \rangle / \sqrt{N} $ remains well-controlled in all cases.


\section{Results} \label{sec:results}

After the introduction of the non-equilibrium setup and the class of initial 
states, we now turn to a discussion of the induced dynamics. First, we discuss 
in Sec.\ \ref{Sec:Universal} the independence of these dynamics of the 
perturbation strength. Then, we discuss two important consequences and present 
specific numerical results, in the context of thermalization (Sec.\ 
\ref{sec:thermalization}) and transport (Sec.\ \ref{Sec:Broad}). 


\subsection{Independence of the perturbation strength} \label{Sec:Universal}

\begin{figure}[tb]
\centering
\includegraphics[width=0.85\columnwidth]{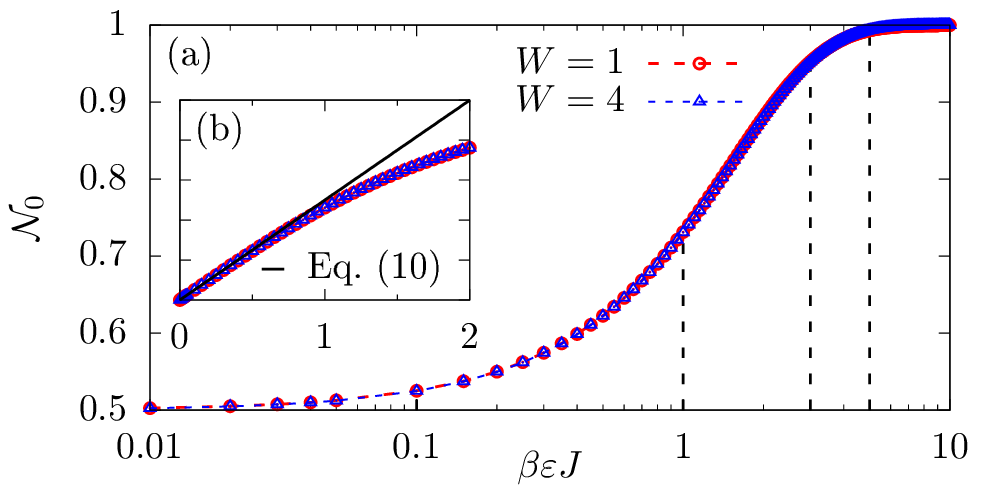}
\caption{(Color online) Initial expectation value ${\cal N}_0 = \langle 
n_{L/2}(0) \rangle$ versus perturbation strength $\varepsilon$ for a high 
temperature $\beta J = 0.01$. Panel (a): Semi-logarithmic ($x$ axis) plot. The 
vertical dashed lines indicate those values of $\varepsilon$ which are used to 
study dynamics, i.e., $\beta\varepsilon J=0.1,1,3,5$. Panel (b): Same data as in 
(a), but now in a linear plot. The 
linear prediction from Eq.\ \eqref{Eq::Linear} is shown. Note that for sufficiently small $\varepsilon$, the response 
is always small compared to the disorder potential $W$. The other parameters 
are $L = 20$ and $\Delta = 1.5$.} \label{Fig::Fig2}
\end{figure}

Let us start by presenting numerical results. As a first step, we study the 
expectation value $\langle n_{L/2}(t) \rangle$, i.e., we measure the 
occupation-number operator $n_l$ at the same lattice site $l = L/2$ which is 
used to prepare the initial state. Before discussing dynamics, it is 
instructive to study the dependence of the initial expectation value ${\cal N}_0 
= \langle n_{L/2}(0) \rangle$ on the strength of the perturbation. In Fig.\ 
\ref{Fig::Fig2}, ${\cal N}_0$ is shown for a high temperature $\beta J = 0.01$ 
up to a perturbation strength $\beta \varepsilon J \leq 10$. One observes that 
${\cal N}_0$ increases linearly for small $\varepsilon$ [see Fig.\ 
\ref{Fig::Fig2} (b)] and eventually saturates for larger $\varepsilon$ to the 
maximum eigenvalue $n_\text{max} = 1$. Furthermore, as expected for such high 
temperatures, ${\cal N}_0$ is independent of the Hamiltonian and therefore the 
curves for $W = 1$ and $W = 4$ in Fig.\ \ref{Fig::Fig2} are practically 
indistinguishable. The vertical dashed lines in Fig.\ \ref{Fig::Fig2} indicate 
those values of $\varepsilon$ which will be used in the following for the study 
of dynamics. Note that these values are chosen in such a way that we cover the 
whole range from states close to equilibrium up to states which are maximally 
perturbed. 

\begin{figure}[tb]
\centering
\includegraphics[width=0.85\columnwidth]{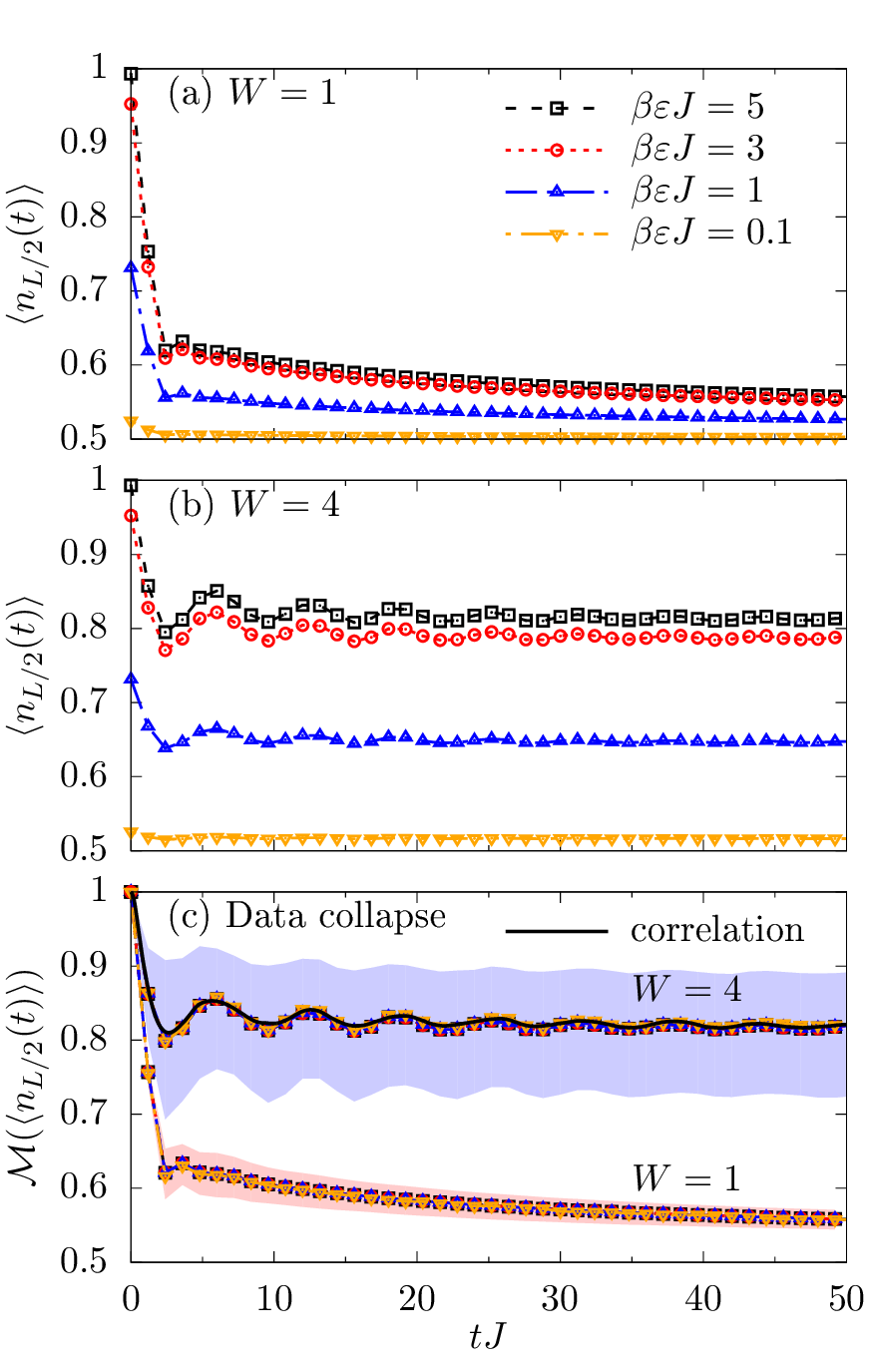}
\caption{(Color online) Dynamical expectation value $\langle n_{L/2}(t)\rangle$ 
for a high temperature $\beta J = 0.01$ and various perturbation strengths 
$\varepsilon$ up to times $t J \leq 50$. Panels (a) and (b) show data for 
disorder strengths $W = 1$ and $W = 4$. Panel (c) shows a collapse of the data 
according to Eq.\ \eqref{Eq::collapse}. The solid line indicates the 
(normalized) equilibrium correlation function $\langle n_{L/2}(t) n_{L/2} 
\rangle_\text{eq}$. The shaded area indicates the statistical fluctuations 
$\overline{\langle n_l(t)\rangle} \pm \Delta\langle n_l(t) \rangle$ 
[cf. Eq.~\eqref{Eq::Error}] due to the 
random potentials. Note that the error of the mean $\Delta \langle n_l(t) 
\rangle/\sqrt{N}$ is significantly smaller. The other parameters are $L = 20$, 
$\Delta = 1.5$, and $N = 300$.} \label{Fig::Fig3}
\end{figure}

Let us now discuss dynamical expectation values. In Figs.\ \ref{Fig::Fig3} (a) 
and (b), $\langle n_{L/2}(t)\rangle$ is shown for a high temperature $\beta J = 
0.01$ and various perturbation strengths $\varepsilon$ for two different 
disorder values $W = 1$ and $W = 4$. Starting with the case $W = 1$, we find a 
quick decay of $\langle n_{L/2}(t)\rangle$ at short time scales $t J \lesssim 
5$, followed by a significantly slower decay towards the long-time value 
$\langle n_{L/2}(t\to\infty)\rangle \sim n_\text{eq} = 1/2$ (although this value is not yet reached at the maximum time 
shown here). On the other hand, for $W = 4$, $\langle n_{L/2}(t)\rangle$ 
exhibits some oscillations which are absent in the case of $W = 1$ (cf.\ Ref.\  
\cite{Prelovsek2017}), and more importantly, we clearly find a long-time value 
$\langle n_{L/2}(t \rightarrow \infty) \rangle \gg n_\text{eq}$. Since the 
initial values $\langle n_{L/2}(0)\rangle$ depend on the specific value of the 
perturbation $\varepsilon$ (cf.\ Fig.\ \ref{Fig::Fig2}), all curves in Figs.\ 
\ref{Fig::Fig3} (a) and (b) naturally differ from each other. However, 
following the approach introduced in Ref.\ \cite{Richter2017}, a simple 
rescaling of the form 
\begin{equation} \label{Eq::collapse}
 \mathcal{M}(\langle n_{l}(t) \rangle) = a \langle n_{l}(t) \rangle + b\ ,
\end{equation}
with \textit{time-independent} coefficients $a$ and $b$, 
\begin{equation}
a = \frac{n_\text{max} - n_\text{eq}}{\langle n_{L/2}(0)
\rangle - n_\text{eq}}\ ; \quad b =  (1 - a) n_\text{eq}\ , 
\end{equation}
leads to a collapse of the data for different $\varepsilon$ onto a single 
curve, as shown in Fig.\ \ref{Fig::Fig3} (c). Thus, independent of the specific 
value of $\varepsilon$, i.e., independent of the initial state being close to 
or far away from equilibrium, the resulting time dependence is universal.
Specifically, due to the projection property $n_l^2 = n_l$, one can write 
\cite{Richter2017}
\begin{equation}\label{Eq::Uni}
\langle n_l(t) \rangle = \frac{n_\text{eq} + (e^{\beta \varepsilon} - 1)\langle 
n_l(t) n_{L/2} \rangle_\text{eq}}{1 + (e^{\beta \varepsilon} - 1)n_\text{eq}}\ , 
\end{equation}
i.e., our non-equilibrium dynamics at high temperatures is always given by 
a correlation function at equilibrium.

Thus, we end up with an intriguing situation: The class of initial states 
$\ket{\psi(0)}$, as introduced in Eq.\ \eqref{Eq::State1}, realizes a dynamics 
where the long-time limit $\langle n_{L/2}(t \rightarrow \infty)\rangle$
clearly depends (i) on the value of $\varepsilon$ and (ii) on the strength of  
the disorder $W$, in particular for $W > W_\text{c}$. On the other hand, the overall  
time dependence of $\langle n_{L/2}(t)\rangle$ is completely independent of 
$\varepsilon$. Remarkably, this result also holds true for rather strong 
disorder $W = 4$, where the ETH is known to be violated. This fact also 
illustrates nicely that typicality of random states is unrelated to the 
validity of the ETH.

The above universality of the time dependence results since the 
occupation-number operators $n_l$ satisfy the projection property $n_l^2 = 
n_l$. However, this particular property is clearly lost if one defines the 
local densities according to a larger unit cell. In fact, already if the unit 
cell contains two sites, then the corresponding local density $d_l = n_{2l-1} + 
n_{2l}$ ($1\leq l \leq L/2$) is not a projection operator any more. Thus, a physically important 
question is: Does a similar type of dynamical universality also emerge in this 
case? If not, the previous discussion would have been about a mathematical 
singularness and not about physical properties of the system.

To answer this question, one can use the fact that all $n_l$ mutually commute. 
As a consequence, the exponential $e^{\beta \varepsilon d_{l'}} = e^{\beta 
\varepsilon n_{2l'-1}} \, e^{\beta \varepsilon n_{2 l'}}$ ($l' = L/4$) can be written as a 
product of two individual exponentials. 
Therefore, using the projection property 
$n_{l'}^2 = n_{l'}$ again, we find
\begin{equation}
e^{\beta \varepsilon d_{l'}} = \Big [1 + (e^{\beta \varepsilon} -1) n_{2{l'}-1} \Big] 
\Big [1 + (e^{\beta \varepsilon} -1) n_{2{l'}} \Big ] \, ,
\end{equation}
which can be multiplied out and, using the abbreviation $g(\beta 
\varepsilon) = e^{\beta \varepsilon} -1$, rewritten as
\begin{equation}
e^{\beta \varepsilon d_{l'}} = 1 + g(\beta \varepsilon) \, d_{l'} + g(\beta 
\varepsilon)^2 \, n_{2{l'}-1} n_{2{l'}} \, .
\end{equation}
For the dynamical expectation value $\langle d_{l}(t) \rangle = 
\text{Tr}[\rho(t) d_{l}]$ with the initial density matrix $\rho = 
e^{\beta \varepsilon d_{l'}} / \text{Tr}[e^{\beta \varepsilon d_{l'}}]$, this relation 
then yields
\begin{equation} \label{2sites}
\langle d_{l}(t) \rangle = \frac{1 + g(\beta \varepsilon) \langle d_{l'} 
d_{l}(t) \rangle_\text{eq} + g(\beta \varepsilon)^2 C(t)}{1 + g(\beta 
\varepsilon) + g(\beta \varepsilon)^2/4} \, ,
\end{equation}
where we have also used the high-temperature averages $\langle d_l 
\rangle_\text{eq} = 1$ for any $l$ and $\langle n_l n_{l'} \rangle_\text{eq} = 
1/4$ for $l \neq l'$. Clearly, this equation is different from Eq.\ 
(\ref{Eq::Uni}) due to the factor $g(\beta \varepsilon)^2$ but, most
importantly, because of the correlation function $C(t) = \langle n_{2{l'}-1} 
n_{2{l'}} \, d_{l} (t) \rangle_\text{eq}$. Thus, in general, the time dependence 
cannot be expected to be independent of the perturbation $\varepsilon$.

\begin{figure}[tb]
\centering
\includegraphics[width=0.85\columnwidth]{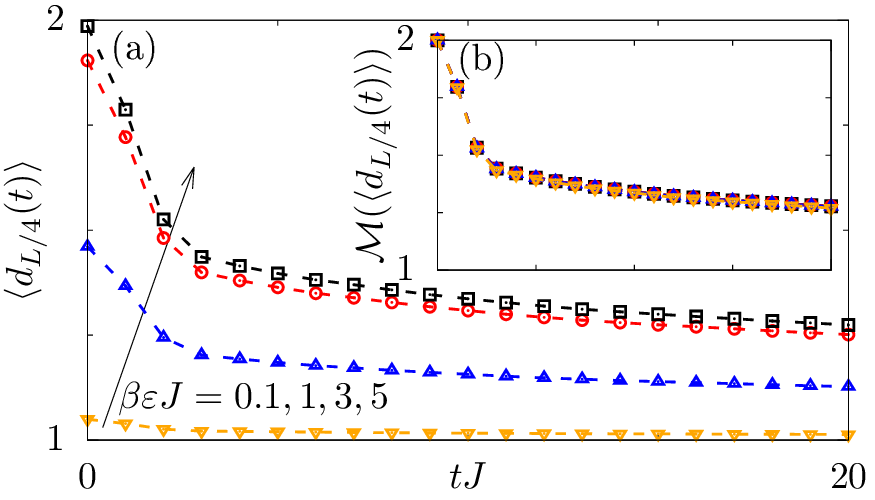}
\caption{(Color online) Dynamical expectation value $\langle d_{L/4}(t)\rangle$
for a high temperature $\beta J = 0.01$ and various perturbation strengths
$\varepsilon$ up to times $t J \leq 20$. Panel (b) shows a collapse of the data 
according to Eq.\ \eqref{Eq::collapse} with $d_\text{eq} = 1$ and $d_\text{max} 
= 2$. The other parameters are $L = 20$, $\Delta = 1.5$, and $W = 
1$.} \label{Fig4}
\end{figure}

However, if we assume that correlation functions for particles and holes behave 
the same,
\begin{equation}
C(t) \overset{!}{=} \langle [1-n_{2{l'}-1}] [1-n_{2{l'}}] [2 - d_{l} (t)] \rangle_\text{eq} \, ,
\end{equation}
we get $C(t) = \langle d_{l'} d_{l}(t) \rangle_\text{eq}/2 -1/4$ and, as a 
consequence, the nominator of Eq.\ (\ref{2sites}) becomes
\begin{equation} \label{2sites_2}
1 + g(\beta \varepsilon) \langle d_{l'} 
d_{l}(t) \rangle_\text{eq} + g(\beta \varepsilon)^2 [\langle d_{l'} 
d_{l}(t) \rangle_\text{eq}/2 -1/4] \, .
\end{equation}
Therefore, in the case of a particle-hole symmetric system, the only time dependence is 
generated by the correlation function $\langle d_{l'} d_{l}(t) 
\rangle_\text{eq}$, even in the case of a two-site unit cell for the definition 
of the local density. This prediction is also confirmed numerically in Fig.\ 
\ref{Fig4}. We note that repeating the calculation for an ever larger unit cell 
yields higher powers in $g(\beta \varepsilon)$.

\begin{figure}[tb]
\centering
\includegraphics[width=0.85\columnwidth]{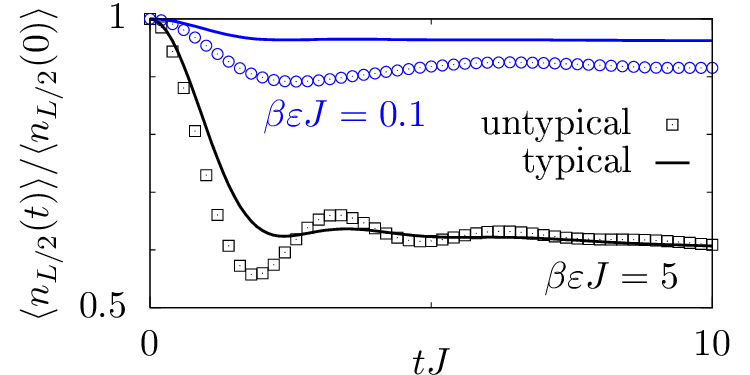}
\caption{(Color online) Comparison between a typical and an untypical state. 
The other parameters are $L = 20$, $\Delta = 1.5$, $W = 1$, and $\beta J = 0.01$.} \label{Fig5}
\end{figure}

As already pointed out, Eq.\ (\ref{Eq::Uni}) as well as Eq.\ (\ref{2sites_2}) 
apply to the overwhelming majority of pure states drawn at random from a 
high-dimensional Hilbert space. But there should be counterexamples, of course. 
One of these counterexamples is a pure state $| \psi(0) \rangle$ which is 
still given by the definition in Eq.\ (\ref{Eq::State1}) but results from a
specific reference state $| \varphi \rangle$ with all coefficients $c_k = 
\text{const.}$ being the same, to which we refer as an untypical state. Such a pure state $| \psi(0) \rangle$ is a valid 
member of the ensemble. However, the probability to draw this state at random 
is certainly tiny: ${\cal O}(2^{-L})$. For this untypical $| \psi(0) \rangle$, 
we show in Fig.\ \ref{Fig5} the time-dependent expectation value $\langle 
n_{L/2}(t) \rangle$ for different values of the perturbation $\varepsilon$ and a 
single set of other model parameters. Compared to Fig.\ \ref{Fig::Fig3}, the 
time dependence is apparently different, and it does change with $\varepsilon$ 
as well.

%


\subsection{Eigenstate thermalization} \label{sec:thermalization}

Let us start with a discussion of the validity of the ETH. To study 
this validity for the Hamiltonian $\mathcal{H}$ in Eq.\ \eqref{Eq::Ham} and 
the occupation-number operator $n_l$, one can introduce the following two 
quantities \cite{Steinigeweg2014}:
\begin{equation} \label{Eq::ETHQ}
\bar{n} = \sum_{i = 1}^{2^L} p_i \bra{i} n_l \ket{i}\ , \quad \Sigma^2 = 
\sum_{i=1}^{2^L} p_i \bra{i} n_l \ket{i}^2 - \bar{n}^2\ , 
\end{equation}
with $\mathcal{H} \ket{i} = E_i \ket{i}$ as well as $p_i \propto 
e^{-(E_i-E)^2/2(\delta E)^2}$ and $\sum_i p_i = 1$. Thus, $\bar{n} = 
\bar{n}(E, \delta E)$ is a weighted average of the diagonal matrix elements 
$n_{ii} = \bra{i} n_l \ket{i}$ in the eigenbasis of $\mathcal{H}$, and most 
sensitive to a (microcanonical) energy region of width $\delta E$ around $E$. 
Likewise, $\Sigma^2 = \Sigma^2(E,\delta E)$ is a weighted variance of the  
$n_{ii}$ in this energy region. Moreover, since these quantities should be 
practically independent of the specific lattice site (if one averages over  
disorder), we just calculate them for a single $l \in [1, L]$. If the ETH 
applies, the diagonal matrix elements $n_{ii}$ should be a smooth function of 
energy in the thermodynamic limit. Consequently, $\Sigma^2$ should become small 
in this case \cite{Beugeling2014, Steinigeweg2013}. While it is certainly 
possible to obtain $\bar{n}$ and $\Sigma^2$ by exact diagonalization of small 
systems, we here also rely on a useful typicality-based approach 
\cite{Steinigeweg2014} to calculate these quantities for larger systems. 
Details on this approach are given in Appendix \ref{sec:ETH_typicality}. 

In Figs.\ \ref{Fig:ETH} (a) and (c), $\bar{n}$ as well as $\bar{n} \pm
\Sigma$ are shown in the energy range $E/J = [-3,3]$ (roughly in the center of 
the spectrum), for the two disorder strengths $W = 1$ and $W = 4$, respectively. 
In both cases, we choose an energy resolution $\delta E / J = 0.5$. Moreover, we compare 
data for $L = 12$ (exact diagonalization) and $L = 20$ (typicality-based 
approach). Starting with the case $W = 1$, we find that $\Sigma$ (i) becomes 
slightly larger for increasing $E$ (cf.\ Ref.\ \cite{Beugeling2014}) and (ii) 
visibly decreases with increasing system size $L$. Although we do not perform a 
concrete finite-size scaling here (see e.g.\ Ref.\ \cite{Beugeling2014} for the 
disorder-free case $W=0$), Fig.\ \ref{Fig:ETH} (a) is consistent with a 
vanishing $\Sigma$ in the thermodynamic limit, i.e., the ETH is fulfilled for 
the small disorder $W = 1$. On the contrary, for $W = 4$, one observes that 
$\Sigma$ is not a function of $E$ and, even more importantly, it does practically 
not scale with $L$ at all. Thus, for $L \to \infty$, $\Sigma$ will be nonzero 
and the ETH is violated for the strong disorder $W = 4$.

The apparent differences between the two cases $W = 1$ and $W = 4$ are also 
clearly visible when studying the cloud of diagonal matrix elements directly. 
In Figs.\ \ref{Fig:ETH} (b) and (d), the matrix elements $n_{ii}$ are shown 
versus the corresponding eigenenergies $E_i$, whereby we focus on a single 
subspace with $L/4$ fermions ($L = 20$) and consider only a
single realization of disorder ($N = 1$). While in the case of $W = 1$ the 
$n_{ii}$ are aligned relatively close to each other, they appear randomly 
distributed for $W = 4$ with enhanced probability at the extrema $n_{ii} = 0$, 
$1$ (see Ref.\ \cite{Roy2018} for similar results).

\begin{figure}[tb]
\centering 
\includegraphics[width=0.85\columnwidth]{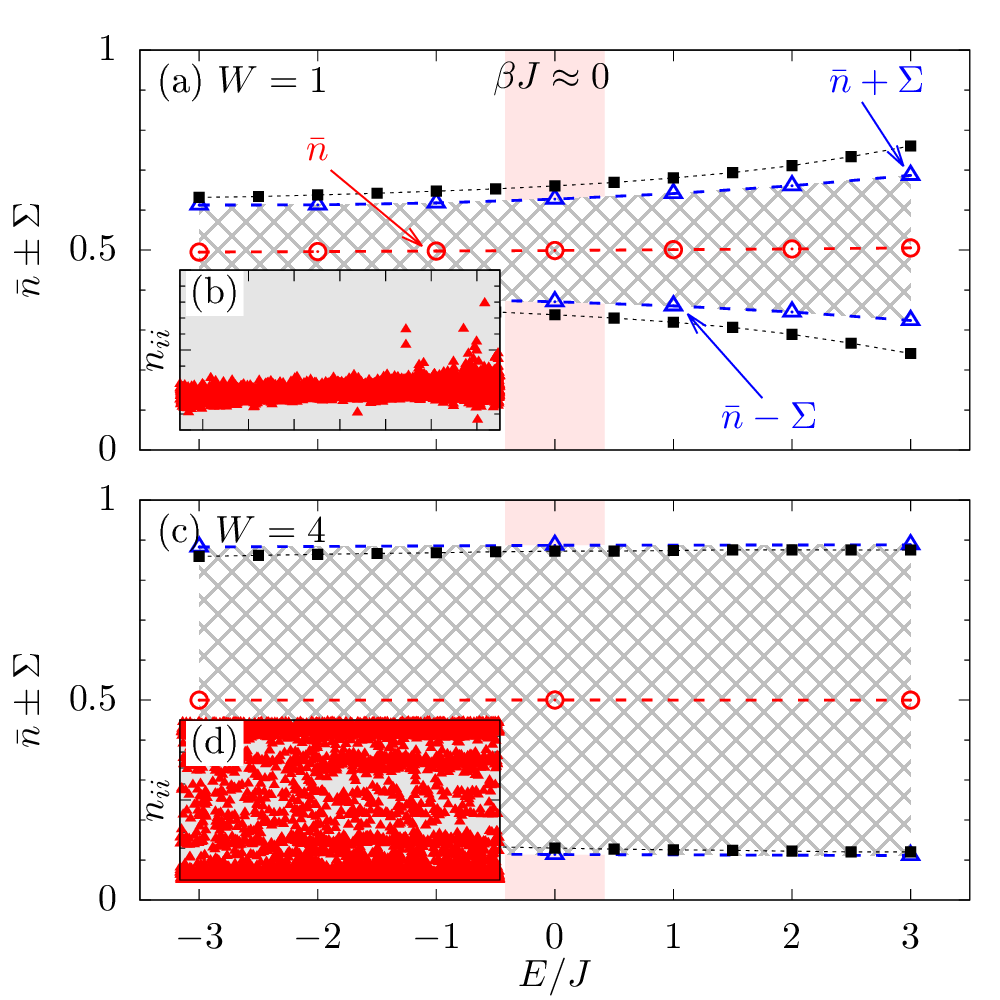}
\caption{(Color online) Panels (a) and (c): Weighted average $\bar{n} \pm 
\Sigma$ in the energy range $-3 \leq E/J \leq 3$ for $W = 1$ and $W = 4$, 
respectively. Filled symbols are exact-diagonalization results for $L = 12$, 
open symbols are obtained by the typicality-based approach for $L = 20$. The 
microcanonical window around $E/J \approx 0$ corresponds to high temperatures 
$\beta J \approx 0$. Panels (b) and (d): Distribution of diagonal matrix elements $n_{ii}$ 
versus eigenenergy $E_i$ in the subspace with $L/4$ fermions 
($L = 20$). The other parameters are $\delta E/J = 0.5$ and 
$\Delta = 1.5$.}
\label{Fig:ETH}
\end{figure}

Let us now establish a relation between the quantities $\bar{n}$ and 
$\Sigma$ and the non-equilibrium dynamics $\langle n_l(t) \rangle$ discussed 
before. In view of Eq.\ \eqref{Eq::Uni}, we find that the long-time value of 
$\langle n_l(t) \rangle$ follows as (see also Appendix \ref{sec:ETH_typicality})
\begin{equation}\label{Eq::LT}
 \langle n_{L/2}(t\to\infty) \rangle = c_1 + c_2(\bar{n}^2 + \Sigma^2)\ , 
\end{equation}
with the two $\beta \epsilon$ dependent coefficients
\begin{equation}
c_1 = \frac{n_\text{eq}}{1 + (e^{\beta\varepsilon} - 1) n_\text{eq}}\ ;\ c_2 = 
\frac{e^{\beta\varepsilon} - 1}{1 + (e^{\beta\varepsilon} - 1) n_\text{eq}}\ .
\end{equation}
One readily sees that for $\Sigma = 0$ (and $\bar{n} = n_\text{eq}$), Eq.\  
\eqref{Eq::LT} reduces to $\langle n_{L/2}(t \to \infty) \rangle = 
n_\text{eq}$, independent of $\beta \epsilon$. Thus, if $\Sigma=0$, 
$\langle n_l(t) \rangle$ relaxes towards the equilibrium 
value $n_\text{eq}$, irrespective of $\beta \epsilon$. In contrast, 
if $\Sigma > 0$, $\langle n_l(t) \rangle$ does not reach its 
equilibrium value at long times and, in particular, this long-time value is 
directly given by the width of the distribution of the diagonal matrix 
elements. Note that in any finite system one expects
\begin{equation}
 \Sigma \geq \Sigma_\text{min} = \frac{1}{\sqrt{4L}}\ , 
\end{equation}
which is a consequence of Eq.\ \eqref{Eq::LT} and 
$\langle n_{L/2}(t\to\infty)\rangle \geq ({\cal N}_0 - n_\text{eq})/L + n_\text{eq}$, 
i.e., the initial $\delta$ peak is eventually distributed over a finite number of lattice sites only. 

\begin{figure}[tb]
\centering
\includegraphics[width=0.85\columnwidth]{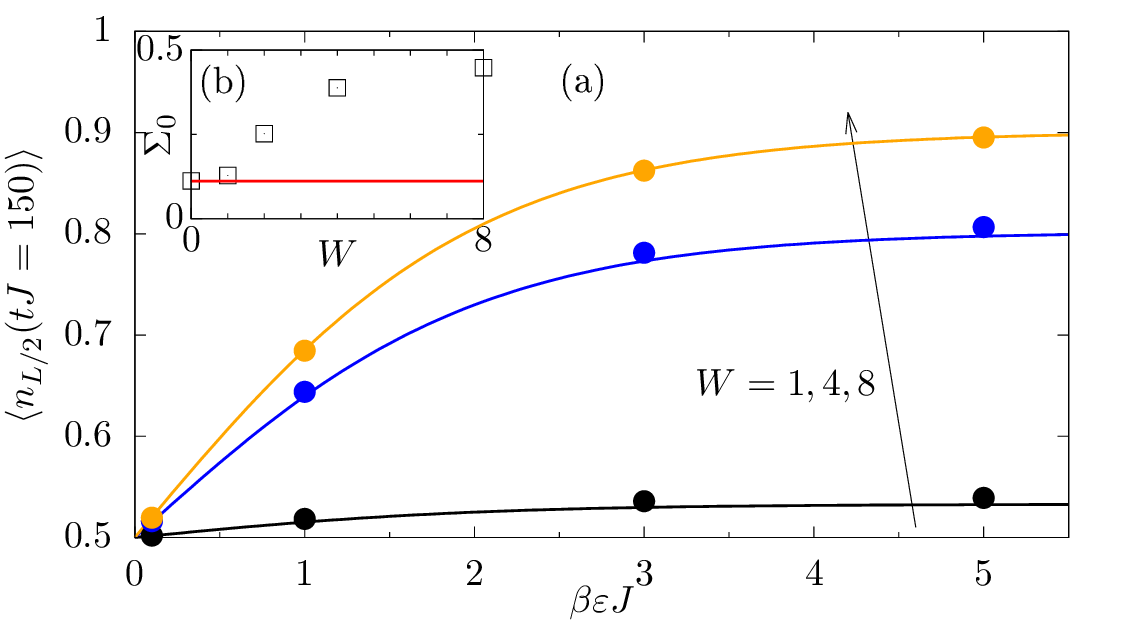}
\caption{(Color online) Panel (a): Numerical illustration of the validity of  
Eq.\ \eqref{Eq::LT}. The data points are extracted from the non-equilibrium 
dynamics $\langle n_{L/2}/(t)\rangle$ at time $t J = 150$, while the solid 
lines are calculated according to Eq.\ \eqref{Eq::LT} with $\Sigma_0 = 
\Sigma(E/J=0)$. In addition to the cases 
$W = 1$ and $W = 4$, we also show data for the case $W=8$ (see
Appendix \ref{sec:LongerTimes}). Panel (b): $\Sigma_0$ versus disorder $W$. 
For comparison, the lower bound $\Sigma_\text{min} = 1/\sqrt{4L}$ is indicated, which follows 
from Eq.\ \eqref{Eq::LT} and $\langle n_{L/2}(t\to\infty)\rangle \geq ({\cal N}_0 - n_\text{eq})/L + n_\text{eq}$ in 
systems of finite size (horizontal line).
In all cases, we have $L=20$ and $\Delta = 1.5$.
} \label{Fig:ETH2}
\end{figure}

It is important to stress that the quantities $\bar{n}(E)$ and $\Sigma(E)$ on the 
r.h.s.\ of Eq.\ \eqref{Eq::LT} have to be chosen from a microcanonical energy 
window $[E - \delta E, E + \delta E]$ corresponding to high temperatures 
$\beta J \approx 0$, in the sense of the equivalence of ensembles. While this 
procedure might not be justified a priori in a disordered system, we 
depict in Fig.\ \ref{Fig:ETH2} (a), the long-time value of $\langle 
n_{L/2}(t) \rangle$, extracted at time $t J = 150$ and for the parameters in Fig.\ 
\ref{Fig::Fig3} (see also Appendix \ref{sec:LongerTimes}). Moreover, we 
compare these data with the prediction in Eq.\ \eqref{Eq::LT}, where $\Sigma_0 = 
\Sigma(E/J = 0)$ should 
correspond to $\beta J \approx 0$. Generally, one observes a 
convincing agreement of the data for all values of $W$ and $\beta \epsilon$ 
shown here. The residual deviations are presumably not only caused by 
statistical fluctuations due to the random potentials, but also by the finite time $t J = 150$ considered. 
(Note that the symbols lie above the solid lines.) Overall, however, Fig.\ 
\ref{Fig:ETH2} (a) confirms Eq.\ \eqref{Eq::LT}, which implies that the validity 
of the ETH is a necessary condition for the thermalization of our class of 
initial states. 

%


\subsection{Broadening of non-equilibrium profiles} \label{Sec:Broad}

So far, we have only considered the single expectation value $\langle 
n_{L/2}(t) \rangle$. Now, we intend to discuss the dynamics of $\langle 
n_{l}(t) \rangle$ for all $1 \leq l \leq L$. First, let us reiterate that the 
states $\ket{\psi(0)}$ realize an initial density profile with a $\delta$ peak 
on top of a homogeneous many-particle background, cf.\ Eq.\ 
\eqref{Eq::Profile}. This $\delta$ peak will gradually broaden over 
time according to the Schr\"odinger equation, and we aim at classifying the 
particular type of broadening. 

To begin with, the real-time and real-space dynamics of $\langle n_l(t) 
\rangle$ can be said to be diffusive, if it fulfills the lattice 
diffusion equation \cite{Michel2005, Steinigeweg2007} 
\begin{equation} \label{Eq::DiffEq}
\frac{\text{d}}{\text{d}t}\langle n_l(t) \rangle = D[\langle n_{l-1}(t)\rangle 
- 2 \langle n_l(t)\rangle + \langle n_{l+1}(t)\rangle]\ ,
\end{equation}
where $D$ is a time-independent diffusion constant. For our initial 
$\delta$ profile, a specific solution of Eq.\ \eqref{Eq::DiffEq} is given in 
terms of a Bessel function \cite{Richter2018}, which (for sufficiently large 
$L$ and long $t$) can be very well approximated by the Gaussian function
\begin{equation}
\langle n_l(t) \rangle - n_\text{eq} \propto \exp \left[ 
-\frac{(l-L/2)^2}{2\sigma^2(t)} \right]\ , 
\end{equation}
with the spatial variance $\sigma^2(t) = 2Dt$. Thus, in case of diffusive 
transport, $\langle n_l(t) \rangle$ must be a Gaussian profile of width 
$\sigma(t) \propto \sqrt{t}$. For any type of transport, the spatial variance 
$\sigma^2(t)$ can be also obtained from $\langle n_l(t) \rangle$ according to 
\cite{Richter2018, Karrasch2014, Steinigeweg2017}
\begin{equation} \label{Eq::Sigma1}
\sigma^2(t) = \sum_{l=1}^L l^2 \, \delta n_l(t) - \left(\sum_{l=1}^L l\ \delta 
n_l(t) \right)^2\ ,
\end{equation}
where $\delta n_l(t) = (\langle n_l(t) \rangle - n_\text{eq})/({\cal N}_0 - 
n_\text{eq})$ is introduced such that $\sum_l \delta n_l(t) = 1$. For our 
initial states and any perturbation $\varepsilon$ (in the high-temperature 
limit $\beta J \approx 0$), it follows that the time-derivative of the spatial 
variance $\sigma^2(t)$ is given by \cite{Steinigeweg2009, 
Steinigeweg2009_2, Karrasch2017, Yan2015}, 
\begin{equation} \label{Eq::Sigma2}
\frac{\text{d}}{\text{d}t} \sigma^2(t) = 2 D(t)\ ,
\end{equation}
where $D(t)$ plays the role of a time-dependent diffusion coefficient and is 
connected to the current autocorrelation function at equilibrium via the generalized 
Einstein relation
\begin{equation} \label{Eq::D}
 D(t) = \frac{1}{\chi}\int_0^t \langle j(t') j \rangle_\text{eq}\ \text{d}t'\ , 
\end{equation}
with the static susceptibility $\chi = 1/4$ for $\beta \rightarrow 0$.

Very often, it is instructive to study density dynamics in momentum space 
as well \cite{Steinigeweg2011, Bera2017}. A Fourier transform of the lattice 
diffusion equation in Eq.\ \eqref{Eq::DiffEq} yields 
\begin{equation} \label{Eq::Dq}
\frac{\text{d}}{\text{dt}} \langle n_q (t) \rangle = -2(1-\cos q) D_q(t) 
\langle n_q (t) \rangle \ , 
\end{equation}
where one additionally allows for a time- and momentum-dependent diffusion 
coefficient $D_q(t)$  \cite{Steinigeweg2011}. As usual, the lattice momentum 
$q$ takes on the discrete values values $q = 2\pi k/L$ with $k = 0,1,\dots, 
L-1$. In the limit of small $q$, and for our non-equilibrium setup, this $D_q(t)$ 
coincides with the $D(t)$ in Eq.\ (\ref{Eq::D}).

The behavior of $D_q(t)$ can be manifold: On the one hand, in the 
short-time limit, $D_q(t)$ is independent of $q$ and scales ballistically as 
$D_q(t) \propto t$. On the other hand, outside this trivial short-time limit, 
$D_q(t)$ can in principle have any dependence on $q$ and $t$. Nevertheless, 
diffusion clearly requires $D_q(t) = \text{const.}$ in a hydrodynamic regime of 
sufficiently small $q$ and long $t$. In contrast,  different types 
of transport like subdiffusion (superdiffusion) can be defined as power-law 
scaling of the form $D_q(t) \propto t^\alpha$ with $\alpha < 0$ ($\alpha > 0$). 
Note that $D_q(t)$ does not distinguish between coexisting transport 
channels \cite{Sirker2009}.

\begin{figure}[tb]
\centering
\includegraphics[width=1\columnwidth]{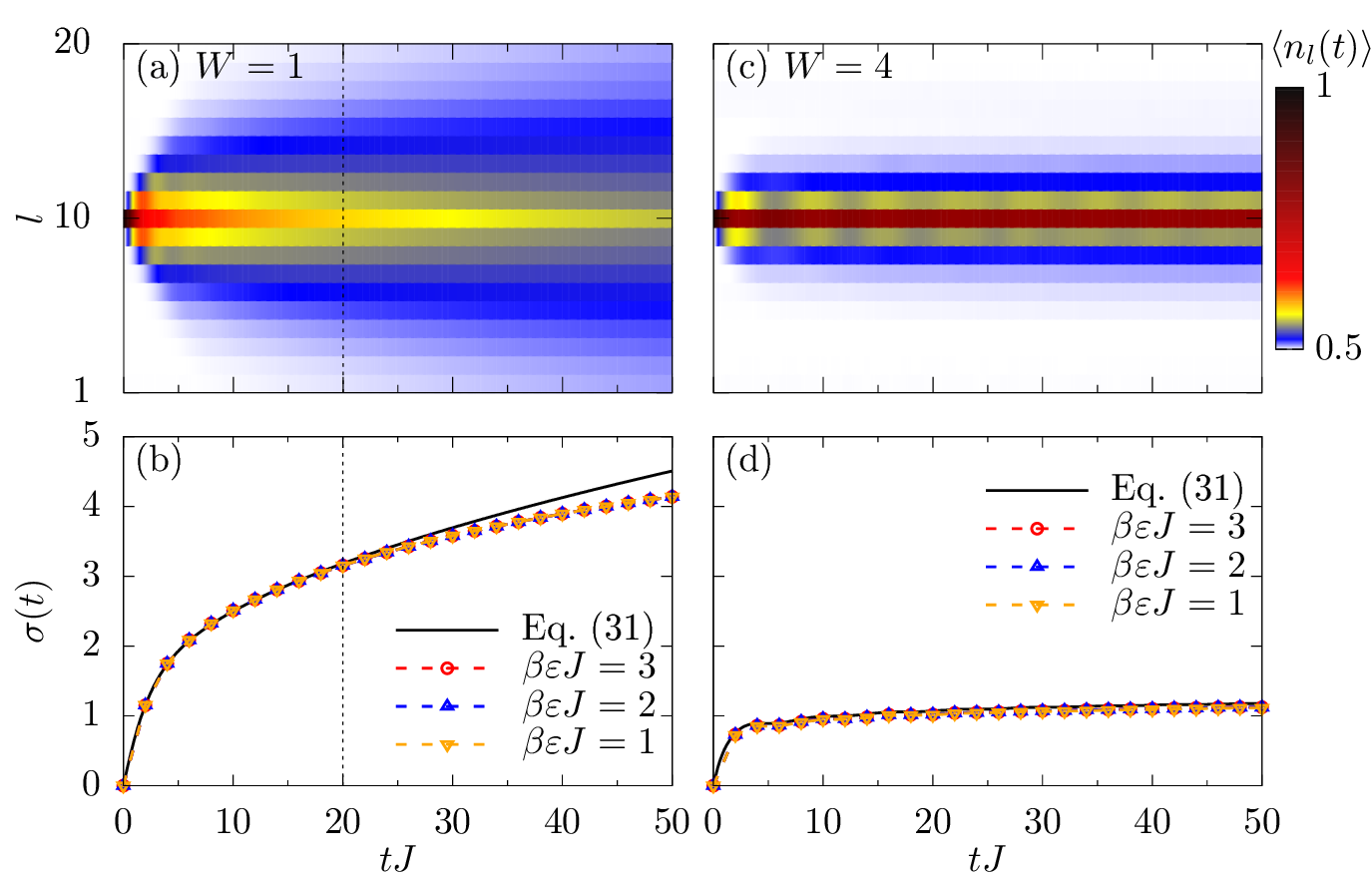}
\caption{(Color online) Panels (a) and (c): Density profile $\langle n_{l}(t) 
\rangle$ for disorder strengths $W = 1$ and $W = 4$, respectively. The initial 
state is prepared according to Eq.\ \eqref{Eq::State1} with $\beta J = 0.01$ 
and $\beta \varepsilon J = 3$. Panels (b) and (d): Corresponding width 
$\sigma(t)$ for various $\beta \varepsilon$, obtained from Eq.\ 
\eqref{Eq::Sigma1}. For comparison, we also depict $\sigma(t)$ from the current 
autocorrelation function, calculated according to Eq.\ \eqref{Eq::Sigma2}. The 
dashed vertical line is a guide to the eye. The other parameters are $L = 20$ and $\Delta 
= 1.5$.} \label{Fig::Fig8}
\end{figure}

Let us now present our numerical results. In Fig.\ \ref{Fig::Fig8} (a), we  
depict the time evolution of the non-equilibrium density profile $\langle n_l(t) 
\rangle$ for a moderate disorder $W = 1$ up to times $t J \leq 50$. On the one 
hand, we can clearly observe the initial $\delta$ profile at $t = 0$. On the 
other hand, this profile broadens relatively quickly and reaches the boundaries 
of the system at $t J \approx 20$ (as indicated by the dashed vertical line). 
In Fig.\ \ref{Fig::Fig8} (b), the corresponding width $\sigma(t)$ of the 
density profile is shown. In agreement with our earlier discussion of the 
universal dynamics in Sec.\ \ref{Sec:Universal}, we find that also $\sigma(t)$ 
is independent of the perturbation $\varepsilon$. Furthermore, according to our 
discussion in the context of Eqs.\ \eqref{Eq::Sigma1} and \eqref{Eq::Sigma2}, 
we compare the profile width to the width obtained from the current 
autocorrelation function. While for short times $t J \lesssim 20$, we find a 
good agreement between both widths, one clearly observes deviations at longer 
times. These deviations can be explained by the fact that a calculation of 
$\sigma^2(t)$ according to Eq.\ \eqref{Eq::Sigma2} is not justified anymore if 
the width of the density profile becomes comparable to the size of the system \cite{Steinigeweg2009}.
It also important to note that a visualization of the data as done here nicely 
illustrates the time scales where finite-size effects due to boundary effects 
become non-negligible.

In Figs.\ \ref{Fig::Fig8} (c) and (d) we show results for the larger disorder $W = 4$.
In contrast to the previous case of $W = 1$, the initial $\delta$ peak broadens 
significantly slower and, even at times $tJ = 50$, the boundary is not reached 
yet. This fact is also reflected by the width $\sigma(t)$ which, after an 
initial increase below $t J \lesssim 5$, saturates to a constant plateau. 
Moreover, in this case, the agreement between calculations of $\sigma(t)$ via
Eq.\ \eqref{Eq::Sigma1} and Eq.\ \eqref{Eq::Sigma2} is excellent for all times 
depicted.

\begin{figure}[tb]
\centering
\includegraphics[width=0.85\columnwidth]{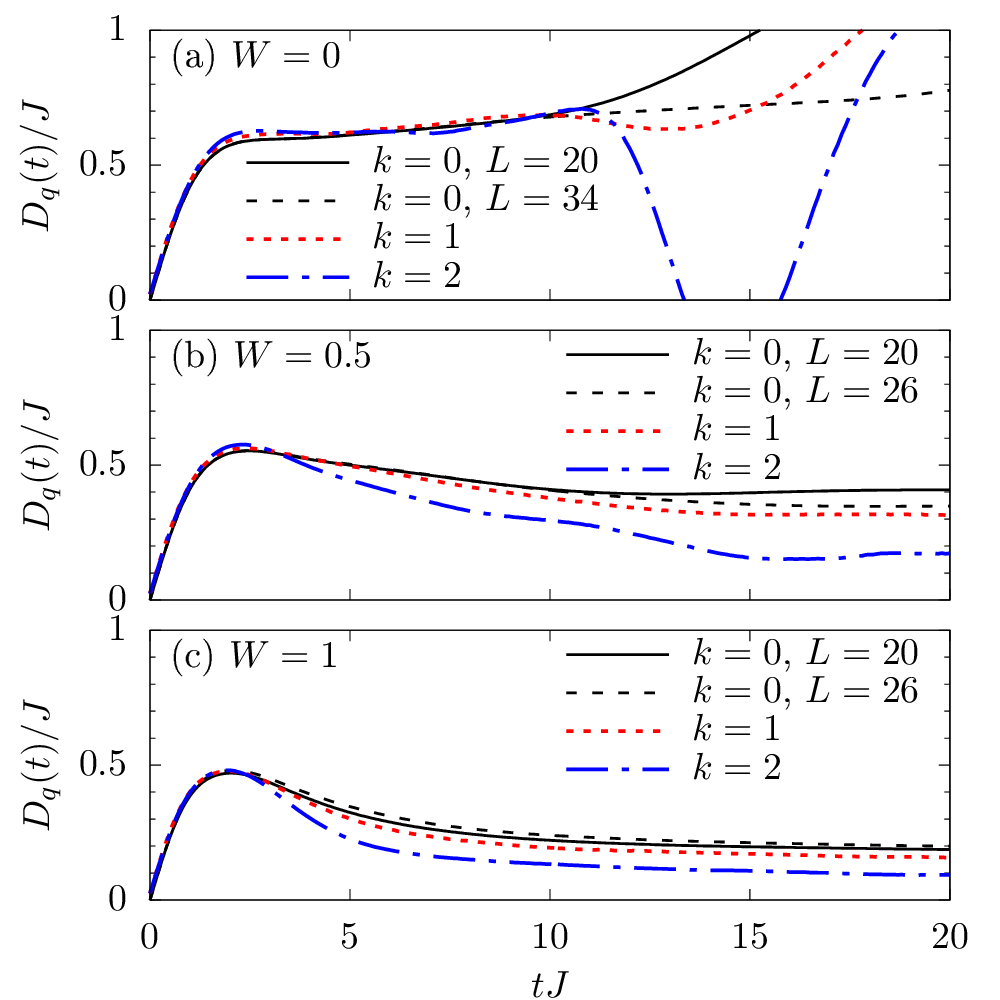}
\caption{(Color online) Time dependence of the generalized diffusion 
coefficient $D_q(t)$ for wave vectors $k = 0$ [Eq.\ \eqref{Eq::D}] and $k = 
1,2$ [Eq.\ \eqref{Eq::Dq}], for three disorders $W = 0$, $0.5$ and $1$, 
respectively. For $k = 0$, data are always shown for two system sizes: (a) $L = 
20$ and $L = 34$; (b) and (c) $L = 20$ and $L = 26$. For $k = 1,2$, we have 
$L = 20$ in all cases. Other parameters: $\Delta = 1.5$.} \label{Fig::Fig9}
\end{figure}

We now turn to studying the broadening of the $\delta$ peak in more 
detail. Specifically, we restrict ourselves to the parameter regime 
of small disorder $W \leq 1$ \cite{Luitz2017, Agarwal2015, Gopalakrishnan2015, 
BarLev2015, Luitz2016, Khait2016}, where sample-to-sample fluctuations are 
still small, cf.\ Fig.\ \ref{Fig::Fig3} (c). In Fig.\ \ref{Fig::Fig9}, the 
diffusion coefficient $D_q(t)$ is depicted for momenta $q/(2\pi/L) = 0$, 
$1$, $2$, times $tJ \leq 20$, and disorder $W = 0$, $0.5$, $1$, i.e., including 
the disorder-free case $W = 0$. For this clean case, we find that $D_q(t)$ is 
approximately constant for times $2 \lesssim tJ \lesssim 10$. Furthermore, at  
these times, we find that $D_q(t)$ coincides for all three momenta $q$ depicted 
\cite{Steinigeweg2011,Richter2018}. Visible differences for longer times are a 
consequence of finite-size effects, as evident when comparing the two $q = 
0$ curves for $L=20$ and $L=34$ \cite{Steinigeweg2014_2}. Hence, for $\Delta = 1.5$, we clearly find 
diffusion in the absence of disorder \cite{Prelovsek2004, Znidaric2011, 
Steinigeweg2011, Karrasch2014, Steinigeweg2017, Ljubotina2017}. In fact, to 
have this well-behaving point of reference, we have chosen $\Delta = 1.5$ 
throughout our paper, in contrast to the vast majority of works on disordered 
systems, which study the isotropic point $\Delta = 1$. 

When switching on small disorder $W > 0$, one clearly observes two changes. 
First, quite counterintuitively, the number of coinciding $q$ is reduced. 
However, at least the two smallest momenta $q/(2\pi/L) = 0$ and $1$ behave 
still the same way. Second, after the initial increase of $D_q(t)$, it 
decreases again. Nevertheless, this decrease then turns into a minor time 
dependence of $D_q(t)$. Hence, at times $10 \lesssim t J \lesssim 20$, no big 
error results when approximating $D_q(t)$ by a constant. Certainly, one might 
be tempted to consider even longer times. But finite-size effects appear at 
such times, as evident when comparing the two $q = 0$ curves for $L=20$ 
and $L=26$ ($\ll 34$). And surely, one might be tempted to analyze the minor 
time dependence in more detail. But such an endeavor is meaningless, as the 
spanned scale at the $y$ axis is much smaller than one order of magnitude.
Consequently, we conclude that our data for small disorder $W = 0.5$ and $1$ are
still consistent with diffusion, while it cannot rule out subdiffusion.

\begin{figure}[t]
\centering
\includegraphics[width=0.85\columnwidth]{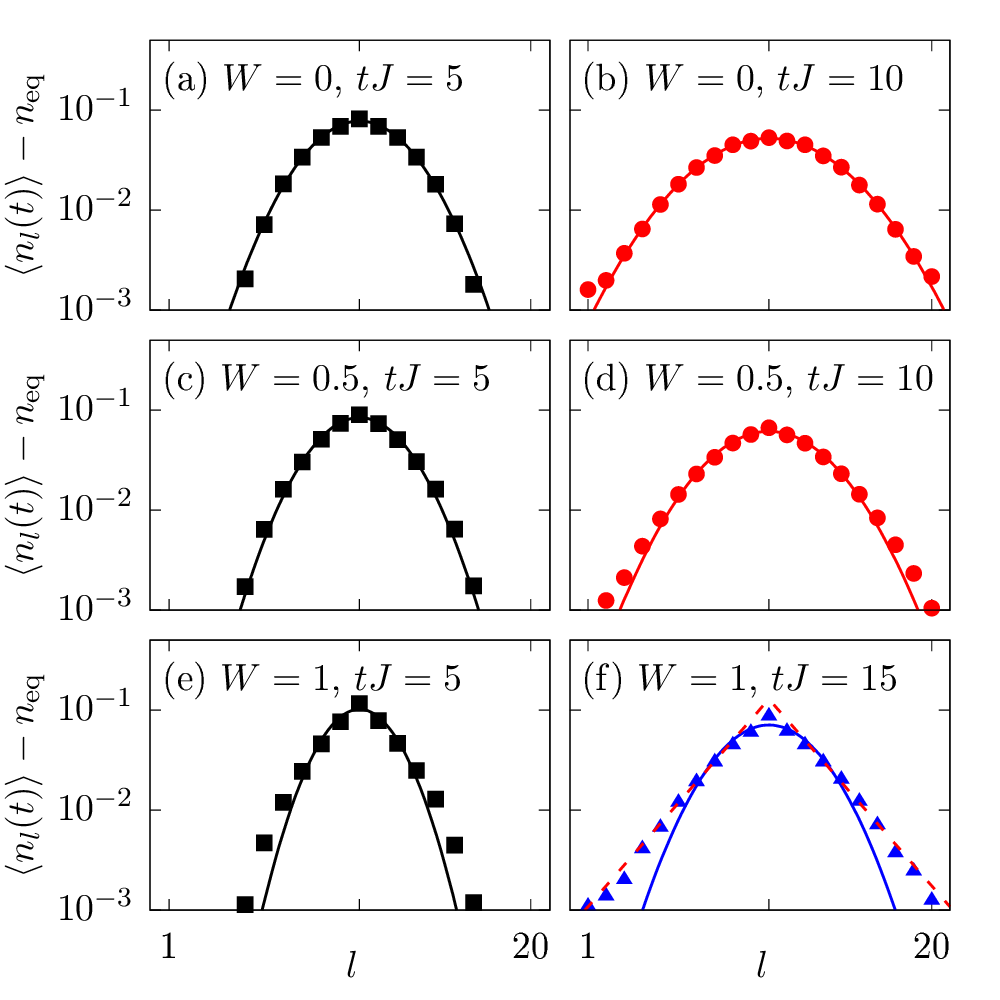}
\caption{(Color online) Density profile $\langle n_l(t) \rangle$ 
for the three disorder strengths (a) $W = 0$, (c) $W = 0.5$, (e) $W = 1$ and a short time $t J 
= 5$, in a semi-logarithmic plot ($y$ axis). For each $W$, a longer time is shown in (b), (d), (f), where boundary 
effects are still negligibly small. In all (a)-(f), Gaussian fits are indicated 
for comparison. In (f), an exponential fit to the outer tails is depicted. 
Other parameters: $\Delta = 1.5$.} 
\label{Fig::Fig10}
\end{figure}

To shed further light onto the differences between the clean case $W = 0$ and 
the disordered cases $W > 0$, we summarize in Fig.\ \ref{Fig::Fig10} the 
site dependence of the density profile $\langle n_l(t) \rangle$ for $W = 0$ 
(top row), $W=0.5$ (middle row), and $W = 1$ (bottom row). Furthermore, we do so 
for a short time $t J = 5$ (left column) and a longer time (right column), 
where boundary effects are still negligibly small. For no or weak disorder, $W = 
0$ and $W = 0.5$, and all times depicted, one can clearly see that the density 
profile $\langle n_l(t) \rangle$ is very well described by Gaussian fits over 
roughly two orders of magnitude. In agreement with our earlier conclusions, 
this pronounced Gaussian form of the density profile provides another strong 
evidence for the existence of diffusion (see also Ref.\ \cite{Steinigeweg2017} for $W = 0$).

\begin{figure}[t]
\centering
\includegraphics[width=0.85\columnwidth]{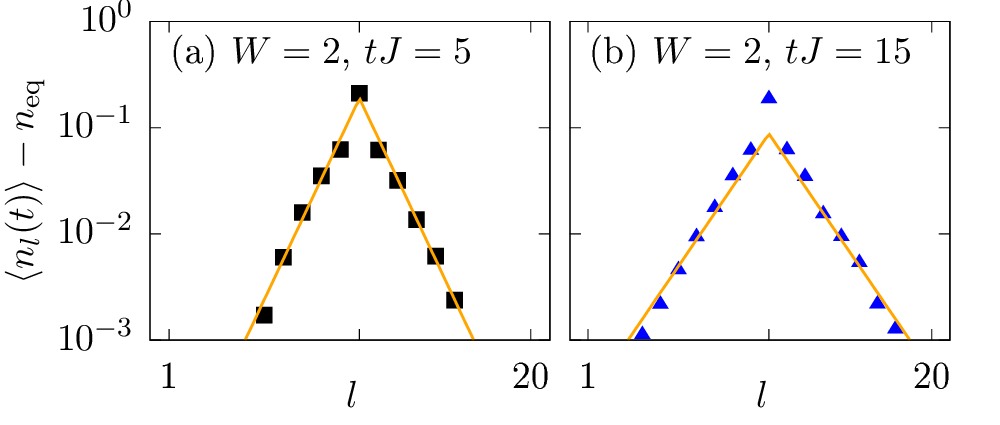}
\caption{(Color online) Same data as in Fig.\ \ref{Fig::Fig10} but now shown 
for $W = 2$.} \label{Fig::Fig11}
\end{figure}

For stronger disorder $W = 1$ and a short time $t J = 5$, the density profile 
can be still described in terms of a Gaussian. It is apparent, however, that 
the agreement is much less convincing. In contrast, for a longer time $t J = 
15$, a Gaussian description clearly fails and is, in particular, not able to 
capture the outer tails of $\langle n_l(t) \rangle$ correctly. Instead, these 
tails appear to be exponential, and the overall density profile has a 
triangular shape in the semi-logarithmic plot used. This shape is 
a signature of non-diffusive dynamics and might be thus consistent with 
subdiffusion in this parameter regime \cite{Znidaric2016}. Despite of larger 
sample-to-sample fluctuations, we find similar results for $W = 2$, see Fig.\ 
\ref{Fig::Fig11}.

\begin{figure}[t]
\centering
\includegraphics[width=0.95\columnwidth]{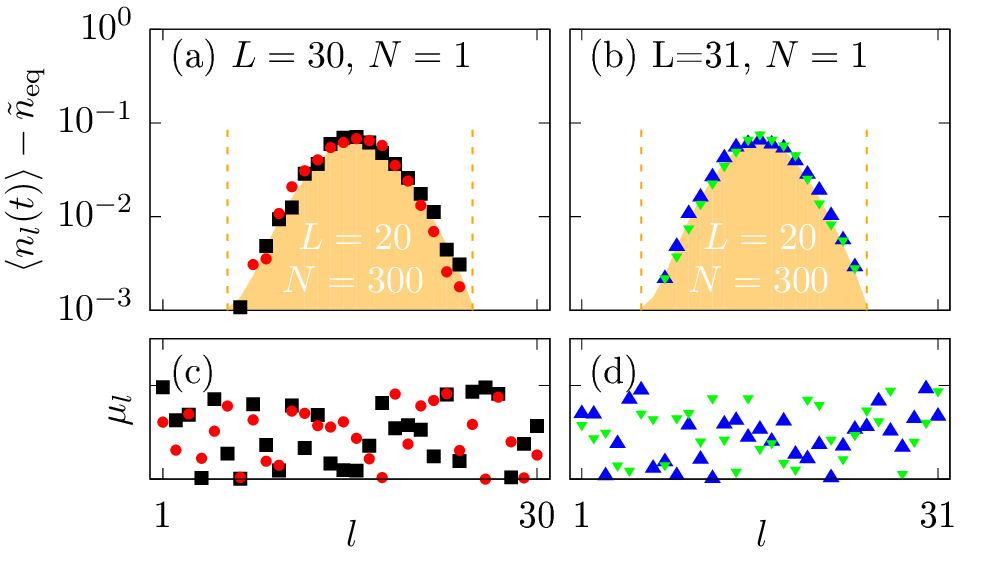}
\caption{(Color online) Density profile $\langle n_l(t) \rangle$ for $W = 
0.5$ and single realizations of the random on-site potential: Two different
realizations for (a) $L = 30$ and (b) $L = 31$ sites. These unaveraged data  
(symbols) are additionally compared to averaged $N = 20$ data (shaded area) 
from Fig.\ \ref{Fig::Fig10} (d) with $t J = 10$. (c), (d) Corresponding 
realizations of the random on-site potential. Note that we restrict 
ourselves to the largest subsector of ($L = 30$) or around $(L=31)$ half filling.}
\label{Fig::Fig12}
\end{figure}

Remarkably, since sample-to-sample fluctuations are small for a small amount 
of disorder, the density profiles $\langle n_l(t) \rangle$ can be accurately 
obtained already from a single realization of the random potential. To 
demonstrate this fact, we repeat the calculation for $W = 0.5$ and $t 
J = 10$ in Fig.\ \ref{Fig::Fig10} (d), but without any averaging over disorder 
configurations. Moreover, we do so for two substantially larger system sizes 
$L=30$ and $31$, where the Hilbert space is huge. As summarized in Fig.\ 
\ref{Fig::Fig12}, the corresponding results agree very well with the averaged 
$N = 20$ results, even in the semi-logarithmic plot used again. This agreement 
also demonstrates that finite-size effects are small on this time scale. Note that 
the calculations for $L \geq 30$ have been carried out only for the 
largest particle subsector, in contrast to all other calculations in this paper. Note 
further that $\tilde{n}_\text{eq} = \langle n_{l\neq L/2}(0)\rangle$
is not strictly identical to $n_\text{eq} = 1/2$ but reads
\begin{equation}
  \tilde{n}_\text{eq} = {{L-2}\choose{L/2 -2}} / {{L-1}\choose{L/2 -1}}\  
 \end{equation}
in the half-filling sector ($\tilde{n}_\text{eq} \approx 0.483$ for $L = 30$).


\section{Conclusion} \label{sec:conclusion}

In summary, we have studied the real-time dynamics of local occupation numbers 
in a one-dimensional model of spinless fermions with a random on-site potential 
for a certain class of initial states. These initial states are thermal (mixed 
or pure) states of the model in the presence of an additional static force, but 
become non-equilibrium states after a sudden removal of this static force. For 
this class and high temperatures, we have shown that the induced dynamics is 
given by a single correlation function at equilibrium, independent of the 
initial expectation values being prepared close to equilibrium (by a weak 
static force) or far away from equilibrium (by a strong static force). 
Remarkably, this type of universality holds true in both, the ergodic phase and 
the many-body localized regime. Moreover, it does not depend on the specific 
choice of a unit cell for the local density.

We have particularly discussed two important consequences. First, the long-time 
expectation value of the local density is uniquely determined by the 
fluctuations of its diagonal matrix elements in the energy eigenbasis. Thus, 
the validity of the eigenstate thermalization hypothesis is not only a 
sufficient but also a necessary condition for thermalization. Second, 
the real-time broadening of density profiles is always given by the current 
autocorrelation function at equilibrium via a generalized Einstein relation. 
In the context of transport, we have discussed the influence of disorder for 
large particle-particle interactions, where normal diffusion is known to occur 
in the disorder-free case. Our results suggest that normal diffusion is stable 
against weak disorder, while they are consistent with anomalous diffusion for 
non-weak disorder.
 
Promising future research directions include the generalization to different 
non-equilibrium scenarios as well as the study of lower temperatures. 

\subsection*{Acknowledgements}

This work has been funded by the Deutsche Forschungsgemeinschaft (DFG) - STE 
2243/3-1. We sincerely thank the members of the DFG Research Unit FOR 2692 for 
fruitful discussions. J.H. has been supported by the US Department of Energy (DOE), 
Office of Science, Basic Energy Sciences (BES), Materials Sciences and Engineering Division.


\appendix

\section{ETH and typicality} \label{sec:ETH_typicality}

For completeness, we describe in this section how to calculate the ETH 
quantities $\bar{n}$ and $\Sigma^2$ in Eq.\ \eqref{Eq::ETHQ} by means of a 
typicality-based approach. While we closely follow the derivations presented 
in Ref.\ \cite{Steinigeweg2014}, we also show that Eq.\ \eqref{Eq::LT} follows 
from these derivations. 

First, we introduce the pure state $\ket{\psi_E}$, 
\begin{equation}
\ket{\psi_E} = C_E \ket{\varphi} \quad; \quad C_E = 
e^{-\frac{(\mathcal{H}-E)^2}{4(\delta E)^2}}\ ,
\end{equation}
where $\ket{\varphi}$ is again a random state drawn according to the unitary 
invariant Haar measure (cf.\ Eq.\ \eqref{Eq::State1} and below) and the 
operator $C_E$ is an energy filter of Gaussian type \cite{Steinigeweg2014,Garnerone2013}, 
$C_E^2 = \exp[-(\mathcal{H}-E)^2/2(\delta E)^2]$. Exploiting the concept of typicality, the 
quantity $\bar{n}(E)$ can then be obtained according to
\begin{equation}\label{Eq::BarN}
\bar{n}(E) \approx \frac{\bra{\psi_E} n_l 
\ket{\psi_E}}{\braket{\psi_E|\psi_E}}\ , 
\end{equation}
where the statistical error due to $\ket{\varphi}$ and the dependence  
on $\delta E$ have been dropped for clarity.
Next, in order to calculate 
$\Sigma(E)$, we define
\begin{align}
\gamma_E(t) &= \frac{ \bra{\psi_E} n_l(t) n_l \ket{\psi_E} 
}{ \braket{\psi_E|\psi_E} }\\
&= \frac{ \bra{\psi_E(t)} n_l \ket{\tilde{\psi}_E(t)} }{ \braket{\psi_E|\psi_E} 
} \label{Eq::Gamma}\ ,
\end{align}
where $\ket{\tilde{\psi}_E} = n_l \ket{\psi_E}$ and $\ket{\psi_E(t)} = 
e^{-i\mathcal{H}t} \ket{\psi_E}$. 
Now, we require that $\gamma_E(t)$ relaxes with time to some value and then 
stays approximately constant. Note that, from a numerical point of view, this 
relaxation should be also sufficiently fast, to make our approach efficient. 
Given this requirement, we find that the long-time average of $\gamma_E(t)$ is 
given by \cite{Steinigeweg2014}
\begin{equation}\label{Eq::Gamma2}
 \bar{\gamma}_E = \frac{1}{t_2 - t_1} \int_{t_1}^{t_2} \text{d}t\ \gamma_E(t) 
\approx \bar{n}(E)^2 + \Sigma(E)^2\ .  
\end{equation}
Thus, it is possible to obtain $\Sigma(E)$ according to 
\begin{equation} \label{Eq::Gamma3}
 \Sigma(E) = \sqrt{\bar{\gamma}_E - \bar{n}(E)^2}\ ,
\end{equation}
with $\bar{n}(E)$ as given in Eq.\ \eqref{Eq::BarN}.

Note that Eq.\ \eqref{Eq::Gamma2} implies the validity of Eq.\ \eqref{Eq::LT} 
in the main text because of two reasons: (i) $\gamma(t)$ is nothing else than 
the equilibrium correlation function $\langle n_l(t) n_l \rangle_\text{eq}$ 
within the approximate microcanonical energy window $[E-\delta E, E+\delta E]$. 
(ii) In the context of Eq.\ \eqref{Eq::Uni}, we have discussed that $\langle 
n_l(t) \rangle \propto \langle n_l(t) n_l \rangle_\text{eq}$ within the 
canonical ensemble at high temperatures $\beta \approx 0$. Thus, in the sense of 
the equivalence of ensembles, there is a direct relation between the long-time 
values of $\gamma(t)$ and $\langle n_l(t) \rangle$, if the microcanonical energy 
$E$ is chosen is such a way that it corresponds to the canonical temperature 
$\beta \approx 0$.

\begin{figure}[t]
\centering
\includegraphics[width=0.8\columnwidth]{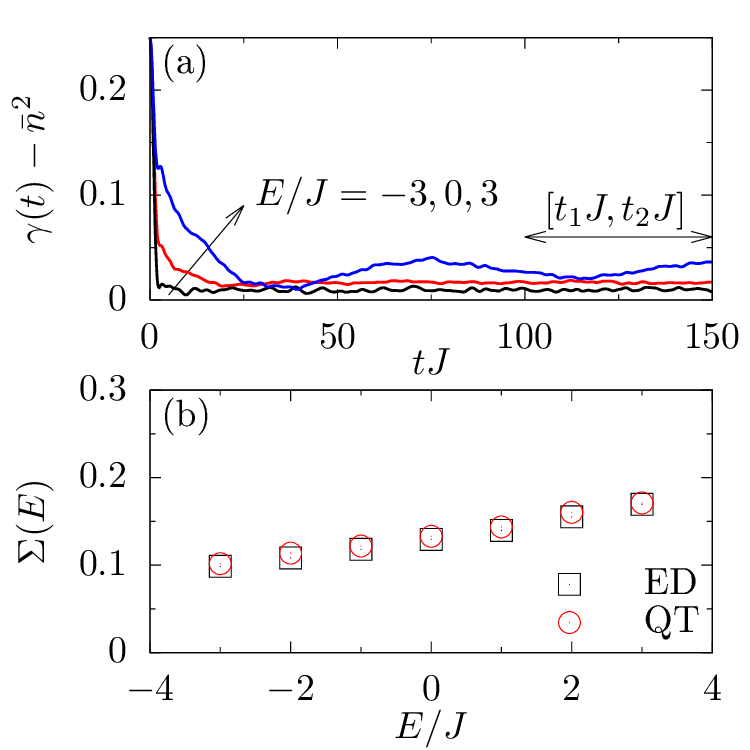}
\caption{(Color online) Panel (a): Quantity  $\gamma(t) - \bar{n}^2$ according 
to Eqs.\ \eqref{Eq::BarN} and \eqref{Eq::Gamma} for $L = 14$ sites, averaged 
over $N=300$ random realizations of the pure state $\ket{\varphi}$. Data are 
shown for energies $E/J = -3,0,3$. Panel (b): Comparison of the width 
$\Sigma(E)$, as obtained from the exact definition in Eq.\ \eqref{Eq::ETHQ}, 
with the width $\Sigma(E)$, as obtained from Eqs.\ \eqref{Eq::Gamma} -- 
\eqref{Eq::Gamma3} with the time interval $[t_1 J,t_2 J] = [100,150]$. The other 
parameters are $\Delta = 1.5$ and $W = 0$.} 
\label{Fig::Fig13}
\end{figure}

To illustrate the accuracy of this pure-state approach, we compare in 
Fig.\ \ref{Fig::Fig13} (b) the width $\Sigma(E)$, as obtained from the exact 
definition in Eq.\ \eqref{Eq::ETHQ}, with the width $\Sigma(E)$, as 
obtained from Eqs.\ \eqref{Eq::Gamma} -- \eqref{Eq::Gamma3} within the time 
interval $[t_1 J,t_2 J] = [100,150]$, cf. Fig.\ \ref{Fig::Fig13} (a). We find a 
very good agreement between both approaches, already for the small system with
$L = 14$ sites. Note that, while we average here over $N = 300$ realizations of 
the pure state $\ket{\varphi}$ to reduce statistical fluctuations, this 
averaging becomes less important for increasing system size.  

\section{Longer times and other values of disorder} \label{sec:LongerTimes}

In the main text, we have mainly focused on the two disorder strengths $W = 1$ 
and $4$ and considered times up to $tJ \leq 50$. For completeness, let us here 
also show data for $W = 2$ and  $8$ as well as longer times. Note that 
these data has been already used in the context of Fig.\ \ref{Fig:ETH2}. 

\begin{figure}[tb]
\centering 
\includegraphics[width=\columnwidth]{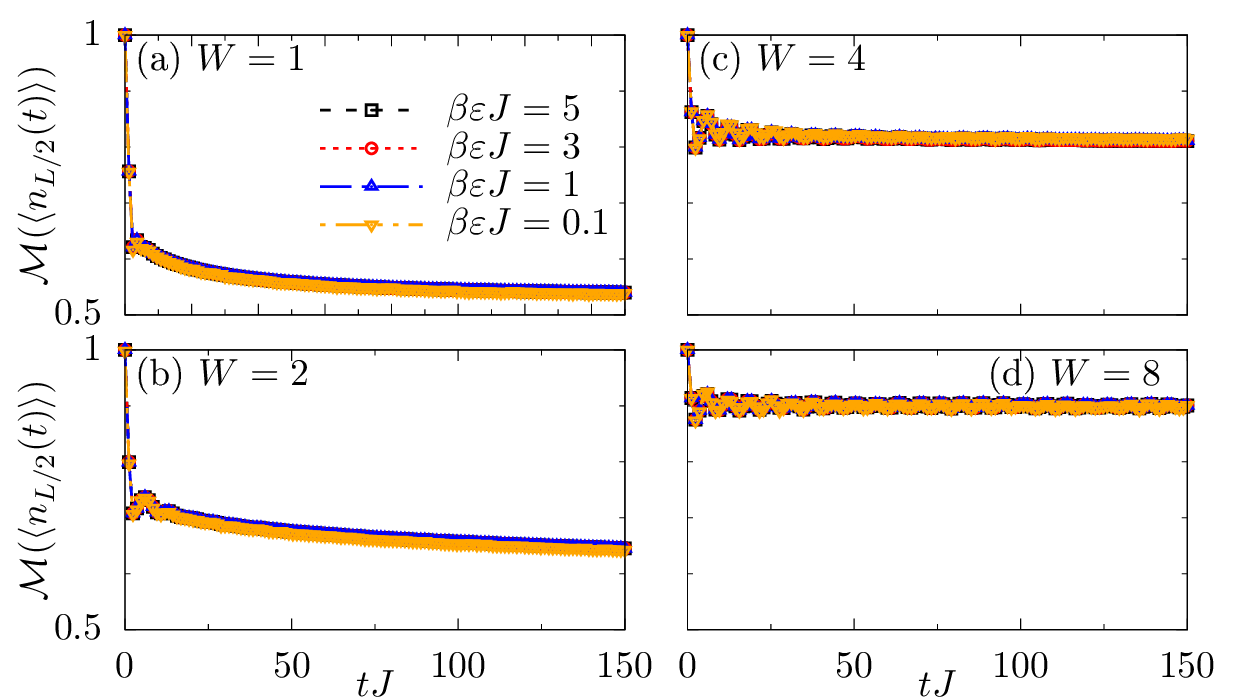}
\caption{(Color online) Data collapse ${\cal M}(\langle n_{L/2}(t) \rangle)$,
cf.\ Eq.\ \eqref{Eq::collapse}, for perturbations $\beta \varepsilon J = 0.1,1,3,5$ and disorder strengths $W = 1,2,4,8$ up to long times 
$tJ \leq 150$. The other parameters are $L = 20$ and $\Delta = 1.5$.} \label{Fig14}
\end{figure}

For the dynamical expectation value $\langle n_{L/2}(t) \rangle$ at the 
single site site $l = L/2$, we depict in Fig.\ \ref{Fig14} the data collapse 
${\cal M}(\langle n_{L/2}(t) \rangle)$ for disorder strengths $W = 1$, $2$, 
$4$, and $8$ up to long times $tJ\leq 150$. For the intermediate disorder $W = 
2$, one finds that even at these long times, $\langle n_{L/2}(t) \rangle$ has 
not yet reached its final value. 

%

\begin{figure}[b]
\centering 
\includegraphics[width=\columnwidth]{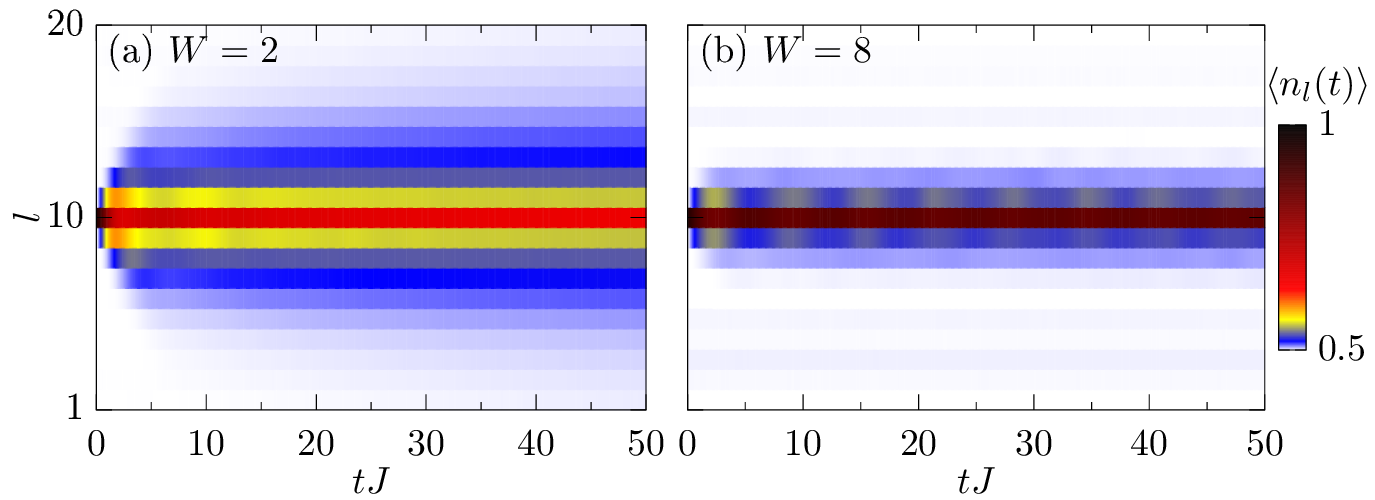}
\caption{(Color online) Density profile $\langle n_{l}(t) \rangle$ for disorder 
strengths $W = 2$ and $W = 8$, respectively. The initial state is prepared 
according to Eq.\ \eqref{Eq::State1} with the parameters $\beta J = 0.01$ and 
$\varepsilon \beta J = 3$. The other parameters are $L = 20$ and $\Delta = 1.5$.}
 \label{Fig15}
\end{figure}

Additionally, we show in Fig.\ \ref{Fig15} the time evolution of the full 
density profile $\langle n_l(t) \rangle$, $0\leq l \leq L$, for the two 
disorder strengths $W = 2$ and $8$, complementary to the data already 
presented in Fig.\ \ref{Fig::Fig8}. While the $\delta$ peak broadens 
for $W = 2$, the dynamics are essentially frozen 
for $W = 8$.

\section{Calculation of correlation functions}

The numerical calculation of equilibrium correlation functions such as
$\langle n_l(t) n_l \rangle_\text{eq}$ or $\langle j(t) j \rangle_\text{eq}$
is an important aspect of our paper. Thus, let us briefly describe how 
these dynamical quantities can be obtained from both, exact 
diagonalization and a typicality-based pure-state approach. In this context, we 
also comment on the class of non-equilibrium pure states $\ket{\psi(0)}$ in 
Eq.\ \eqref{Eq::State1} in more detail.

For simplicity, we focus on the equilibrium correlation function $\langle 
n_l(t) n_l \rangle_\text{eq}$. For the use of exact diagonalization, this 
correlation function is conveniently written in terms of the eigenstates 
$\ket{a}$ and the corresponding eigenvalues $E_a$ of the Hamiltonian 
$\mathcal{H}$,
\begin{align}
\langle n_l(t) n_l \rangle_\text{eq} &= \frac{\text{Tr} [e^{-\beta \mathcal{H}} 
\, n_l(t) n_l ]}{\mathcal{Z}_\text{eq}} \label{Eq:App:Korr} \\
&=  \sum_{a,b=1}^{2^L}  \frac{e^{-\beta E_a}}{\mathcal{Z}_\text{eq}} \, | \! 
\bra{a} n_l \ket{b} \!|^2 \, e^{i(E_a - E_b)t}\ ,  \label{Eq:App:Korr2}
\end{align}
with $\mathcal{Z}_\text{eq} = \sum_a e^{-\beta E_a}$. Because of the random 
potentials, exact diagonalization becomes relatively costly
for systems with $L \sim 14$ sites already since (i) translational 
invariance is broken and (ii) also an averaging over a sufficiently large 
number of instances of these random potentials is required.

Using the concept of quantum typicality, on the other hand, the trace in Eq.\ 
\eqref{Eq:App:Korr} can be replaced by a scalar product with a single pure 
state $\ket{\varphi}$, which is drawn at random according to the Haar measure, 
i.e., according to Eq.\ (\ref{phi}) with Gaussian distributed coefficients 
$c_k$. Thus, by introducing the two auxiliary pure states \cite{Elsayed2013, 
Steinigeweg2014_2, iitaka2003} 
\begin{align}
\ket{\phi(t,\beta)} &= e^{-i \mathcal{H} t} \, n_l \, e^{-\beta \mathcal{H}/2} 
\, \ket{\varphi}\ , \\
\ket{\varphi(t,\beta)} &= e^{-i \mathcal{H} t} \, e^{-\beta \mathcal{H}/2}  \,
\ket{\varphi}\ ,
\end{align}
we can rewrite the correlation function $\langle n_l(t) n_l 
\rangle_\text{eq}$ in the form
\begin{equation}\label{Eq:App:QT}
\langle n_l(t) n_l \rangle_\text{eq} = \frac{\bra{\varphi(t,\beta)} n_l 
\ket{\phi(t,\beta)}}{\braket{\varphi(0,\beta) | \varphi(0,\beta)}} + 
f(\ket{\varphi})\ ,
\end{equation}
where the statistical error $f(\ket{\varphi})$ decreases exponentially fast 
with increasing the Hilbert-space dimension.

The numerical evaluation of Eq.\ \eqref{Eq:App:QT} can be done by 
the forward propagation (in real and imaginary time) of the two pure states 
$\ket{\varphi(t,\beta)}$ and $\ket{\phi(t,\beta)}$. As mentioned in the 
main body of the text, the involved operators $n_l$ and $\mathcal{H}$ exhibit a 
sparse-matrix representation such that these propagations can be implemented 
memory efficient, which particularly allows for a treatment of significantly 
larger systems compared to exact diagonalization. Note that Eq.\ 
\eqref{Eq:App:QT}, as well as Eq.\ \eqref{Eq:App:Korr2}, can be used for the 
current autocorrelation function $\langle j(t) j \rangle_\text{eq}$ as well, 
simply by replacing $n_l$ by $j$.

Eventually, let us comment on the non-equilibrium pure states $\ket{\psi(t)}$ 
and its expectation value $\langle n_l (t) \rangle$, as discussed in the main 
part of this paper. In the limit of small $\beta$, we have shown
that (i) the non-equilibrium dynamics is independent of the perturbation 
$\varepsilon$ and that (ii) these dynamics are also identical to the 
equilibrium correlation function. Compared to the above typicality approach 
based on Eq.\ \eqref{Eq:App:QT}, one needs to propagate \textit{one} pure 
state in (real and imaginary time) only, which is certainly a numerical 
advancement. It should be noted, however, that the properties (i) and (ii) are 
not expected to hold for lower temperatures, at least in general.
%


\end{document}